\documentclass[%
 reprint,
amsmath,amssymb,
 aps,
superscriptaddress]{revtex4-1}

\newcommand\Tstrut{\rule[4ex]{0pt}{0pt}}        
\newcommand\Bstrut{\rule[-3ex]{0pt}{0pt}}   
\usepackage{book tabs}
\usepackage{graphicx}
\usepackage{dcolumn}
\usepackage{bm}
\usepackage{relsize}
\usepackage{float}
\usepackage{adjustbox}
\usepackage{color}
\usepackage[normalem]{ulem}
\usepackage{appendix}
\usepackage{tensor}

\begin{document}

\preprint{APS/123-QED}

\title{Screening and degenerate kinetic self-acceleration from the nonlinear freedom of reconstructed Horndeski theories}

\author{Joe~Kennedy}
\affiliation{Institute for Astronomy, University of Edinburgh, Royal Observatory, \\ Blackford Hill, Edinburgh, EH9 3HJ, U.K.}
\author{Lucas~Lombriser}
\affiliation{D\'{e}partement de Physique Th\'{e}orique,
Universit\'{e} de Gen\`{e}ve,
24 quai Ernest Ansermet,
1211 Gen\`{e}ve 4,
Switzerland}
\author{Andy~Taylor}
\affiliation{Institute for Astronomy, University of Edinburgh, Royal Observatory, \\ Blackford Hill, Edinburgh, EH9 3HJ, U.K.}

\date{\today}

\begin{abstract}
We have previously presented a reconstruction of Horndeski scalar-tensor theories from linear cosmological observables. 
It includes free nonlinear terms which can be added onto the reconstructed covariant theory without affecting the background and linear dynamics. 
After discussing the uniqueness of these correction terms, we apply this nonlinear freedom to a range of different applications.
First we demonstrate how the correction terms can be configured to endow the reconstructed models with screening mechanisms such as the chameleon, k-mouflage and Vainshtein effects.
A further implication is the existence of classes of Horndeski models that are degenerate with standard cosmology to an arbitrary level in the cosmological perturbations.
Particularly interesting examples are kinetically self-accelerating models that mimic the dynamics of the cosmological constant to an arbitrary degree in perturbations. %
Finally, we develop the reconstruction method further to the level of higher-order effective field theory, which under the restriction to a luminal propagation speed of gravitational waves introduces two new free functions per order. These functions determine the corresponding correction terms in the linearly reconstructed action at the same order.
Our results enable the connection of linear cosmological constraints on generalised modifications of gravity and dark energy with the nonlinear regime and astrophysical probes for a more global interpretation of the wealth of forthcoming cosmological survey data.

\end{abstract}

\pacs{Valid PACS appear here}
\maketitle

%%%%% INTRODUCTION %%%%%
\section{Introduction} \label{sec:intro}
%%%%%%%%%%%%%%%%%%%%%%%%

The observation of the late-time accelerated expansion of the Universe~\cite{Perlmutter:1998np, Riess:1998cb} has led to a large number of theoretical models that attempt to explain it.
To date, the $\Lambda$ Cold Dark Matter ($\Lambda$CDM) model, consisting of a cosmological constant $\Lambda$ and dark matter treated as a cold pressureless fluid, remains the most successful among them~\cite{Aghanim:2018eyx}.
Despite its simplicity there remain plenty of open questions.
One of the most pressing issues is the large contribution to $\Lambda$ that should arise from the quantum corrections to the various matter fields in the Universe~\cite{Martin:2012bt, Weinberg1989}.
Taken with the fact that the fundamental nature of dark matter also remains a mystery, these problems have encouraged a great deal of model-building beyond $\Lambda$CDM.
Many of these models involve an additional scalar degree of freedom that may drive the acceleration even in the absence of a cosmological constant~\cite{Clifton:2011jh, Joyce:2014kja, Joyce:2016vqv}. The scalar field can be thought of as an additional exotic contribution to the matter sector or the low-energy effective description of a modification to General Relativity (GR) which acts on cosmological scales.
Scalar fields typically arise through a symmetry breaking mechanism from a UV-complete theory. The Higgs field is an example of this and its presence in the Standard Model of particle physics provides motivation to study the effects of scalar fields on gravitational dynamics. 

Incorporating a scalar field into GR is not a trivial task. Higher derivatives can easily enter the field equations of motion leading to extra propagating degrees of freedom and an unbounded Hamiltonian. This is a consequence of the Ostrogradsky theorem~\cite{Woodard:2006nt}.
In 1974 Horndeski
%wrote down
identified
the unique scalar-tensor theory in four dimensions that leads to at most second-order equations of motion and thus avoids the Ostrogradsky ghost~\cite{Horndeski:1974wa}. The theory was later re-discovered by generalising Galileons to curved spacetime~\cite{Deffayet:2011gz, Kobayashi2011}.
Note that it is possible to have stable higher-order theories~\cite{Gleyzes:2014dya, Langlois:2015cwa}, which shall however not be considered in this work.

A useful approach to a unified treatment of various dark energy and modified gravity models is provided by the effective field theory (EFT) of dark energy.
The formalism was originally developed in the context of inflation~\cite{Weinberg:2008hq, Cheung:2007st} before its application to late-time cosmology~\cite{Creminelli:2008wc, Park:2010cw, Bloomfield:2011np, Gubitosi2012, Bloomfield:2012ff, Gleyzes:2013ooa, Bloomfield:2013efa, Tsujikawa:2014mba, Gleyzes:2014rba, Bellini:2014fua, Lagos:2016wyv}.
It features a systematic order-by-order expansion in the cosmological perturbations and proves to be a useful tool for the unified description of
the cosmological effects of Horndeski theory.

The trade-off for generality is that this EFT formalism
is restricted by definition to certain length scales, usually just the cosmological background and linear perturbations. Recently there has been some work in extending the expansion to higher-order perturbations~\cite{Cusin:2017mzw, Frusciante:2017nfr}. An alternative approach is to start from the full covariant action. The loss of generality in this approach is now traded for the applicability on a much broader range of length scales, allowing nonlinear effects such as screening to be studied.
We have recently presented~\cite{Kennedy:2017sof, Kennedy:2018gtx} a reconstruction from the EFT of dark energy on the level of the background and linear perturbations to the class of Horndeski theories that give rise to the particular set of given EFT functions. With this covariant action it becomes feasible to generally connect the nonlinear regime to that of the background and linear scales. This link shall be the focus of this paper.

More precisely, within the reconstructed theory of Refs.~\cite{Kennedy:2017sof,Kennedy:2018gtx} there are correction terms that account for the nonlinear freedom that exists between Horndeski theories that are degenerate at the level of the background and linear perturbations. Specification of these correction terms allows one to move between linearly degenerate theories.
We first discuss the uniqueness of the correction terms in the reconstructed theory.
Applying the recent constraint on the equality between the speeds of light and of gravitational waves~\cite{Monitor:2017mdv} we show that the number of free functions that are present at higher order in the EFT of dark energy is significantly reduced to two per order in perturbation theory.  
This then implies that the nonlinear freedom is uniquely specified by the nonlinear correction terms.
It is worth noting that out of the four new EFT functions found in Ref.~\cite{Bellini:2015wfa} at second order in the cosmological perturbations of Horndeski theory, the two functions dominating in the sub-horizon regime vanish for a luminal speed of gravity, and the impact of our nonlinear correction terms on the weakly nonlinear regime of structure formation remains to be examined in detail.

As an initial demonstration of the implications of the correction terms, we show how this nonlinear freedom can be used to endow a reconstructed theory with a screening mechanism.
Due to the tight Solar-System constraints on deviations from GR~\cite{Will:2014kxa} it is necessary for a large-scale modification of GR to employ a screening mechanism that suppresses the effects of a fifth force on small scales.
These screening mechanisms fall into one of three categories~\cite{Joyce:2016vqv}: those that screen through deep gravitational potentials
such as the chameleon~\cite{Khoury:2003aq} or symmetron mechanisms~\cite{Hinterbichler:2010es}, screening through first derivatives of the potentials such as k-mouflage models~\cite{Babichev:2009ee} or screening through second derivatives as for the Vainshtein mechanism~\cite{Vainshtein:1972sx}. 
     
A simple scaling method was developed in Refs.~\cite{McManus:2016kxu, McManus:2017itv} to determine whether a given theory possesses an Einstein gravity limit or not. We present an application of this scaling method to the reconstructed theory and demonstrate with three examples that there is enough freedom in the nonlinear regime of a reconstructed theory to obtain, in principle, any of these three screening mechanisms.

A further interesting consequence that arises when considering theories built from the correction terms is that it is simple to construct theories that are indistinguishable from $\Lambda$CDM to arbitrary level in cosmological perturbations. Only observations in the nonlinear regime can be used to distinguish them from $\Lambda$CDM. Such degenerate theories may be built from kinetic terms alone without including a cosmological constant, hence providing a kinetic self-acceleration effect. 

Finally, we present a reconstruction from the nonlinear EFT back to the space of manifestly covariant theories.
This follows a similar structure to the background and linear reconstruction and in principle provides a method for obtaining a Horndeski theory reconstructed from a range of different length scales from the background to the nonlinear regime.      

The paper is organised as follows.
In Sec.~\ref{sec:background} we briefly review Horndeski scalar-tensor gravity, the EFT formalism, the reconstruction method from linear EFT to Horndeski gravity and the nonlinear freedom available for the reconstructed theories. The uniqueness of the nonlinear correction terms in the reconstructed action is examined in Sec.~\ref{sec:uniqueness}.
Sec.~\ref{sec:screening} briefly reviews the scaling method and discusses how the nonlinear freedom in the reconstructed scalar-tensor theories can be used to implement screening effects due to large gravitational potentials and large first or second derivatives of the potential.
In Sec.~\ref{sec:kineticSA} we discuss how the nonlinear freedom can be used to construct models that accelerate the cosmic expansion without a cosmological constant with a suitable choice of kinetic terms,
yet are degenerate with standard cosmology at the background level or even to arbitrary level of perturbations. 
The derivation of a third-order reconstruction is presented in Sec.~\ref{sec:higherorderreconstruction} along with a discussion of the extension to $n$-th order. 
Finally, we provide conclusions on the results in Sec.~\ref{sec:conclusions}.

%%%%%% BACKGROUND %%%%%%
\section{Reconstructed scalar-tensor theories} \label{sec:background}
%%%%%%%%%%%%%%%%%%%%%%%%
%
For the benefit of the unfamiliar reader we shall briefly review  Horndeski gravity in Sec.~\ref{sec:horndeski} before discussing the reconstruction from the EFT of dark energy and modified gravity to manifestly covariant theories in Sec.~\ref{sec:lineft}. Sec.~\ref{sec:nonlinear} then examines the nonlinear freedom in this reconstruction.
The free nonlinear correction terms available will then be applied to screening in Sec.~\ref{sec:screening}, to the formulation of degenerate kinetic self-acceleration effects in Sec.~\ref{sec:kineticSA} and finally to the connection to higher-order EFT in Sec.~\ref{sec:higherorderreconstruction}.

%----- HORNDESKI ------%
\subsection{Horndeski gravity} \label{sec:horndeski}
%----------------------%

The most general scalar-tensor theory in four dimensions that yields at most second-order equations of motion is given by the Horndeski action \cite{Horndeski:1974wa, Deffayet:2011gz, Kobayashi2011}
\begin{equation}
S=\sum_{i=2}^{5} \int d^{4} x \sqrt{-g} \, \mathcal{L}_{i} \,,
\label{eq:Horndeski}
\end{equation}
where the Lagrangian densities $\mathcal{L}_{i}$ are defined as
\begin{eqnarray}
\mathcal{L}_{2} & \equiv & G_{2}(\phi,X) \,, \label{eq:HorndeskiL2} \\
\mathcal{L}_{3} & \equiv & G_{3}(\phi, X)\Box \phi \,, \label{eq:HorndeskiL3} \\
\mathcal{L}_{4} & \equiv & G_{4}(\phi, X)R \nonumber\\
 & & -2G_{4X}(\phi, X) \left[(\Box \phi)^{2}-(\nabla^{\mu}\nabla^{\nu}\phi)(\nabla_{\mu}\nabla_{\nu}\phi)    \right] \,, \label{eq:g4} \\
\mathcal{L}_{5} & \equiv & G_{5}(\phi, X)G_{\mu \nu}\nabla^{\mu}\nabla^{\nu}\phi \nonumber\\
 & & +\frac{1}{3}G_{5X}(\phi, X) \left[(\Box \phi)^{3} -3(\Box \phi) (\nabla_{\mu}\nabla_{\nu}\phi)(\nabla^{\mu}\nabla^{\nu}\phi) \right. \nonumber\\
 & & \left. +2(\nabla_{\mu}\nabla_{\nu}\phi)(\nabla^{\sigma}\nabla^{\nu}\phi)(\nabla_{\sigma}\nabla^{\mu}\phi)   \right] \, , \label{eq:g5}
\end{eqnarray}
with $X \equiv g^{\mu\nu}\partial_{\mu}\phi\partial_{\nu}\phi$. A restriction to the class of Horndeski theories with luminal speed of gravity simplifies the action~\eqref{eq:Horndeski} considerably to~\cite{McManus:2016kxu}
\begin{eqnarray}
\mathcal{L}_{2} & \equiv & G_{2}(\phi,X) \,, \label{eq:HorndeskiL2at0} \\
\mathcal{L}_{3} & \equiv & G_{3}(\phi, X)\Box \phi \,, \label{eq:HorndeskiL3at0} \\
\mathcal{L}_{4} & \equiv & G_{4}(\phi)R  \,, \label{eq:g4at0}
\end{eqnarray}
where $\mathcal{L}_{5}$ can be set to zero.
By varying this reduced Horndeski action in Eqs.~\eqref{eq:HorndeskiL2at0} to \eqref{eq:g4at0} with respect to the metric and the scalar field, one obtains the metric and scalar field equations~\cite{Kobayashi2011,McManus:2016kxu}. They will be needed solely in Sec.~\ref{sec:screening} and the explicit expressions are given in the appendix.

%------ LIN-EFT -------%
\subsection{Reconstruction from linear effective field theory} \label{sec:lineft}
%----------------------%

The effects of Horndeski theory on the cosmological background evolution and the linear perturbations can be described in a convenient manner by adopting the EFT of dark energy \cite{Gubitosi2012, Gleyzes:2013ooa, Gleyzes:2014rba, Bloomfield:2012ff, Bloomfield:2013efa, Tsujikawa:2014mba}.
The relevant action is constructed with the usual spirit of EFT by writing down every operator, in this case the cosmological perturbations, which is consistent with the symmetries imposed on the theory. Time diffeomorphism symmetry is broken in the EFT of dark energy and so every operator which remains invariant under spatial diffeomorphisms is employed.
The scalar field is then the pseudo Nambu-Goldstone boson of broken time translational symmetry.

At the level of the background and linear perturbations the EFT action~\cite{Gubitosi2012,Bloomfield:2012ff} in the notation of Ref.~\cite{Lombriser:2014ira} is given by
\begin{align}
S = & \: S^{(0,1)}+S^{(2)}+S_{M}[g_{\mu\nu},\Psi_m] \,,
\label{eftlag}\\
S^{(0,1)} = & \: \frac{M_{*}^{2}}{2}\int d^{4}x \sqrt{-g} \left[ \Omega(t) R -2\Lambda(t)-\Gamma(t)\delta g^{00} \right] \,,
\label{eq:s01} \\
S^{(2)} = & \int d^{4}x \sqrt{-g} \left[ \frac{1}{2}M^{4}_2(t)(\delta g^{00})^2-\frac{1}{2}\bar{M}^{3}_{1}(t) \delta K \delta g^{00}\right. \nonumber\\
 & \left.-\bar{M}^{2}_{2}(t) \left( \delta K^2-\delta  K^{\mu\nu} \delta K_{\mu\nu} - \frac{1}{2} \delta R^{(3)}\delta g^{00} \right) \right] \,,
\label{eq:s2}
\end{align}
The set $\left\{\Omega(t),\Lambda(t),\Gamma(t),M_{2}^{4}(t),\bar{M}_{1}^{3}(t),\bar{M}_{2}^{2}(t)\right\}$ of the free time-dependent coefficients can be derived for a particular choice of the Horndeski functions $G_{i}$ \cite{Gubitosi2012, Gleyzes:2013ooa}. 
There are alternative bases for the EFT coefficients. A frequently adopted set was introduced by Ref.~\cite{Bellini:2014fua}, which is related to the coefficients in Eqs.~\eqref{eq:s01} and \eqref{eq:s2} by a linear transformation
\begin{eqnarray}
\alpha_{M} & \equiv & \frac{M_{*}^{2}\Omega^{\prime}+2(\bar{M}_{2}^{2})^{\prime}}{M_{*}^{2}\Omega+2\bar{M}_{2}^{2}} \, , \label{Am} \\
\alpha_{B} & \equiv & \frac{M_{*}^{2}H\Omega^{\prime}+\bar{M}_{1}^{3}}{2H\left( M_{*}^{2}\Omega+2\bar{M}_{2}^{2}     \right)} \, , \\
\alpha_{K} & \equiv & \frac{M_{*}^{2}\Gamma+4M_{2}^{4}}{H^{2}\left(M_{*}^{2}\Omega+2\bar{M}_{2}^{2} \right)} \, , \\
\alpha_{T} & \equiv & -\frac{2\bar{M}_{2}^{2}}{M_{*}^{2}\Omega+2\bar{M}_{2}^{2}} \label{At} \,,
\end{eqnarray}
where primes denote derivatives with respect to $\textnormal{ln}\, a$. Furthermore $\alpha_{M}$ denotes the Planck mass evolution rate, $\alpha_{B}$ is related to the coupling between the metric and the scalar field, $\alpha_{K}$ arises as a coefficient of the kinetic term for the scalar field, and $\alpha_{T}$ is the deviation of the speed of gravitational waves from that of light, now determined to be vanishing at late times~\cite{Monitor:2017mdv} (also see Refs.~\cite{Lombriser:2015sxa,Lombriser:2016yzn} for forecasted implications). A further set of EFT functions was recently introduced in Ref.~\cite{2019JCAP...01..041L} with $\alpha_{T}=0$ to avoid stability issues associated with the previous EFT bases.  

Different models such as the cubic Galileon~\cite{Chow:2009fm, PhysRevD.80.121301}, quintessence~\cite{Caldwell:1997ii} and k-essence~\cite{PhysRevD.62.023511, ArmendarizPicon:2000ah} in general give different functional forms for this set~\cite{Gubitosi2012}. In particular, Horndeski theories with luminal speed of gravity imply $\bar{M}_{2}^{2}(t)=0$.
The EFT functions can then directly be related to effective descriptions of a modified Poisson equation and gravitational slip~\cite{Lombriser:2014ira, Gleyzes:2014rba, Lombriser2015, Uzan:2006mf, Amendola:2007rr, Caldwell:2007cw, Hu:2007pj, Zhang:2007nk}
that are probed by cosmological observations~\cite{Ade:2015rim}.
 
In order to link observational constraints on these effective modifications to theoretical constraints on fundamental theories it is useful to formulate a mapping from EFT back to the space of physical covariant theories.
Such a reconstruction was developed in Refs.~\cite{Kennedy:2017sof, Kennedy:2018gtx}. It determines the class of Horndeski theories reconstructed from a given set of EFT coefficients that are degenerate at the level of the cosmological background and the linear perturbations.
Specifically, the reconstruction is given by   
\begin{align}
G_{2}(\phi, X) = & -M_{*}^{2}U(\phi) - \frac{1}{2}M_{*}^{2} Z(\phi)X+a_{2}(\phi)X^{2} \nonumber\\
 &+\Delta G_{2} \,,
\label{eq:G2recon} \\
G_{3}(\phi,X) = & \: b_{0}(\phi)+b_{1}(\phi)X+\Delta G_{3} \,,
\label{eq:G3recon} \\
G_{4}(\phi, X) = & \: \frac{1}{2}M_{*}^{2}F(\phi)+c_{1}(\phi)X+\Delta G_{4} \,,
\label{eq:G4recon} \\
G_{5}(\phi, X)= & \: \Delta G_{5} \,,
\label{eq:G5recon}
\end{align}
where each term in the reconstruction such as $U(\phi)$ and $Z(\phi)$ is dependent on a particular combination of EFT functions. For completeness, the full list of expressions is provided in Table~\ref{tab:solution}.
The $\Delta G_i$ functions denote nonlinear correction terms
that characterize the degenerate class of Horndeski theories.
In particular, the correction terms can be used to move between different theories that only differ at the nonlinear level.

%%% BEGIN TABLE %%%
\begin{table*}[t]
\centering
\begin{tabular}{|c|c|} 
\hline
\multicolumn{2}{|c|}{$U(\phi) = \Lambda + \frac{\Gamma}{2} - \frac{M_{2}^{4}}{2M_{*}^{2}} - \frac{9H\bar{M}_{1}^{3}}{8M_{*}^{2}} - \frac{(\bar{M}_{1}^{3})^{\prime}}{8}+\frac{M_{*}^{2}(\bar{M}^{2}_{2})^{\prime\prime}}{4}+\frac{7(\bar{M}_{2}^{2})^{\prime}H}{4}+\bar{M}_{2}^{2}H^{\prime}+\frac{9H^{2}\bar{M}_{2}^{2}}{2M_{*}^{2}}$} \Tstrut \Bstrut \\ 
\hline
\multicolumn{2}{|c|}{$Z(\phi) = \frac{\Gamma}{M_{*}^{4}} - \frac{2M_{2}^{4}}{M_{*}^{6}} - \frac{3H\bar{M}_{1}^{3}}{2M_{*}^{6}} + \frac{(\bar{M}_{1}^{3})^{\prime}}{2M_{*}^{4}}-\frac{(\bar{M}_{2}^{2})^{\prime\prime}}{M_{*}^{2}}-\frac{H(\bar{M}_{2}^{2})^{\prime}}{M_{*}^{4}}-\frac{4H^{\prime}\bar{M}_{2}^{2}}{M_{*}^{4}}$} \Tstrut \Bstrut  \\ \hline 
\multicolumn{2}{|c|}
{$a_{2}(\phi)=\frac{M_{2}^{4}}{2M_{*}^{8}}+\frac{(\bar{M}^{3}_{1})^{\prime}}{8M_{*}^{6}}-\frac{3H\bar{M}_{1}^{3}}{8M_{*}^{8}}-\frac{(\bar{M}_{2}^{2})^{\prime\prime}}{4M_{*}^{4}}+\frac{H(\bar{M}_{2}^{2})^{\prime}}{4M_{*}^{6}}+\frac{H^{\prime}\bar{M}_{2}^{2}}{M_{*}^{6}}-\frac{3H^{2}\bar{M}_{2}^{2}}{2M_{*}^{8}}$} \Bstrut \Tstrut \\ 
\hline                   
\hspace{1.9cm}$ b_{0}(\phi)=0 $ \hspace{1.9cm}  &  $b_{1}(\phi)=\frac{2H\bar{M}_{2}^{2}}{M_{*}^{6}}-\frac{(\bar{M}_{2}^{2})^{\prime}}{M_{*}^{4}}+\frac{\bar{M}_{1}^{3}}{2M_{*}^{6}} $  \Bstrut \Tstrut  \\ 
\hline 
$F(\phi)=\Omega+\frac{\bar{M}_{2}^{2}}{M_{*}^{2}}$ & $c_1(\phi)=\frac{\bar{M}_{2}^{2}}{2M_{*}^{4}}$ \Bstrut \Tstrut \\ 
\hline
\end{tabular}
\caption{The Horndeski functions $G_i(\phi,X)$ reconstructed from the effective field theory of dark energy at the level of the cosmological background evolution and linear perturbations. The primes indicate a derivative with respect to $\phi$. See Ref.~\cite{Kennedy:2017sof} for the derivation.} 
\label{tab:solution}
\end{table*}
%%% END TABLE %%%

%----- NONLINEAR ------%
\subsection{Nonlinear freedom} \label{sec:nonlinear}
%----------------------%

Under the assumption of luminal speed of gravity~\cite{Monitor:2017mdv} we shall show in Sec.~\ref{sec:uniqueness} that the unique nonlinear correction terms in the reconstructed theory are specified by 
%The nonlinear correction terms in Eqs.~\eqref{eq:G2recon} to \eqref{eq:G5recon} are specified by
%
\begin{equation}
    \Delta G_{2,3} = \sum_{n>2} \xi^{{\scriptscriptstyle(2,3)}}_{n}(\phi)\left(1+\frac{X}{M_{*}^{4}}\right)^{n} \, ,
    \label{DeltaG23}
\end{equation}
where $\Delta G_{4,5}=0$ and $\xi^{{\scriptscriptstyle(i)}}_{n}(\phi)$ are free functions of the scalar field, reflecting the large degree of freedom that exists on nonlinear scales without affecting linear scales.  
These terms arise from noting that in the unitary gauge with the foliation $\phi=t M_{*}^{2}$ the kinetic term of the scalar field becomes $X=\small(-1+\delta g^{00}\small)M_{*}^{4}$. Eq.~\eqref{DeltaG23} is therefore an expansion in $\small(\delta g^{00}\small)^{n}$.  
%
%we discuss their uniqueness in Sec.~\ref{sec:uniqueness}.
%Note here that applying the GW170817 constraint of luminal speed of implies that $\Delta G_{4}=\Delta G_{5}=0$ and the only nonlinear freedom is in $G_{2}$ and $G_{3}$, which will be an important consideration for the discussion in Sec.~\ref{sec:uniqueness}. 

The freedom in the correction term~\eqref{DeltaG23} may be exploited to endow the reconstructed theories with some desired nonlinear features without affecting linear theory.
In particular, $\xi^{{\scriptscriptstyle(i)}}_{n}(\phi)$ can be designed to implement a screening mechanism (Sec.~\ref{sec:screening}) or even to hide a kinetic self-acceleration effect of the cosmic background expansion to an arbitrary level of nonlinear perturbations (Sec.~\ref{sec:kineticSA}).

%%%%%% UNIQUENESS %%%%%%
\section{Uniqueness of the $\Delta G_{i}$ corrections} \label{sec:uniqueness}
%%%%%%%%%%%%%%%%%%%%%%%%

Due to the importance of the $\Delta G_{i}$ nonlinear correction terms for the applications of interest in Secs.~\ref{sec:screening}, \ref{sec:kineticSA} and \ref{sec:higherorderreconstruction} we shall first investigate to what extent these terms are the unique
corrections to the reconstructed Horndeski action in Eqs.~\eqref{eq:G2recon} to \eqref{eq:G5recon}.

Recall that the correction terms in~\eqref{DeltaG23} were inferred from the requirement
that in covariant language $\delta g^{00}=1+X/M_{*}^{4}$.
Successive powers of $1+X/M_{*}^{4}$ therefore yield corrections that do not affect lower-order perturbations, in particular, the background or linear theory.
However, there are of course other operators which can be added to the EFT which will not affect the background and linear dynamics such as $\delta K^{3}$ and $\small(\delta R^{(3)}\small)^{3}$.
In principle a term such as $\delta K^{3}$ could be added to the EFT action, which would affect the dynamics of the second-order perturbations. Note however that for the same reason that $\delta K^{2}$ only appears in combination with $\delta K_{\mu\nu}\delta K^{\mu\nu}$ after $\mathcal{L}_{4}$ is written in the unitary gauge and expanded in the perturbations, it is not possible to simply add $\delta K^{3}$ as there are no terms in the Horndeski action that give rise to this term alone.
More specifically, on the cosmological background $K_{\mu\nu}=Hh_{\mu\nu}$, the perturbation $\delta K=K-3H$ must appear in the combination 
\begin{equation}
    K^{3}-3KK_{\mu\nu}K^{\mu\nu}+2K_{\mu\nu}K^{\mu\sigma}\tensor{K}{^{\nu}_{\sigma}} \, ,
\end{equation}
which gives rise to a number of nonlinear operators in the EFT action involving $\delta K_{\mu\nu}$~\cite{Cusin:2017mzw, Frusciante:2017nfr}.
The only term in the Horndeski action that gives rise to such a combination is in $\mathcal{L}_{5}$.
Following the spirit of EFT one may add these nonlinear operators because they are consistent with the symmetries that we have imposed, but the theory which is underlying such a combination generally violates the
luminal speed of gravity constraint~\cite{McManus:2016kxu} such that we will omit these terms.
By use of the Gauss-Codazzi relation 
\begin{equation}
    R^{(3)}=R-K_{\mu\nu}K^{\mu\nu}+K^{2}-2\nabla_{\nu}\left(n^{\nu}\nabla_{\mu}n^{\mu}-n^{\mu}\nabla_{\mu}n^{\nu} \right) \, ,
\end{equation}
relating the 3-dimensional Ricci scalar $R^{(3)}$ to the 4-dimensional Ricci scalar $R$ and $K_{\mu\nu}$, one can furthermore see that adding on higher powers of $R^{(3)}$ to the EFT in a similar manner will inevitably introduce higher powers of $\delta K$, and the previous argument applies. The same logic also requires $\Delta G_{4}$ and $\Delta G_{5}$ to vanish and the nonlinear freedom is now completely specified by Eq.~\eqref{DeltaG23}. 

An alternative perspective on this argument
is to consider a covariant form of the extrinsic curvature tensor, or for simplicity its trace
\begin{equation}
    K=-\nabla_{\mu}\left( \frac{\partial_{\mu}\phi}{\sqrt{-X}}\right) \, .
\end{equation}
By expressing the denominator in terms of the metric perturbations, Taylor expanding and performing the replacement of $\delta g^{00}$ with $1+X/M_{*}^{2}$, one obtains in schematic form
\begin{equation}
    K=\Box \phi+ F(X,\nabla_{\mu}\phi,\nabla_{\mu} X) \, ,
    \label{covexpansionofK}
\end{equation}
\\
where $F(X,\nabla_{\mu}\phi,\nabla_{\mu} X)$ is some complicated function of the scalar field and derivatives of the scalar field obtained after the expansion, the precise form of which is not relevant to the discussion.
Taking higher powers of $\delta K$ and making use of Eq.~\eqref{covexpansionofK} will lead to terms such as $(\Box \phi)^{n}$.
Such expressions belong either to Horndeski models with non-luminal speed of gravitational waves or beyond-Horndeski theories.
Reversing the logic, it is necessary to start from such a model in order to obtain a nonlinear correction involving a higher power of $\delta K$. 
Therefore, any correction terms to the EFT of dark energy that make use of the operators $(\delta K)^n$ with $n \geq 2$ and $R^{(3)}$ will reconstruct a theory that has a non-vanishing $G_{4X}$ or $G_{5}$ term or a beyond-Horndeski model.

For Horndeski models with luminal speed of gravity, the only nonlinear operators that appear at $n$-th order are therefore
\begin{equation}
    \left(\delta g^{00}\right)^{n} \, \, , \, \, 
    \left(\delta g^{00}\right)^{n-1}\delta K \, .
\end{equation}
which adds two new independent EFT functions per order in the perturbations.
More explicitly, the $n$-th order contribution to the EFT action with $n \geq 3$ is given by
\begin{eqnarray}
    \delta \mathcal{S}^{(n)} = \int d^{4}x
\sqrt{-g} \sum_{i=3}^{n} & & \left[ \bar{M}_{i}^{4}(t)\left(\delta g^{00}\right)^{i} \right. \nonumber\\
  & & \left. + \bar{M}_{i}^{3}(t)\left(\delta g^{00}\right)^{i-1}\delta K \right] \, ,
\label{nthorderEFT}
\end{eqnarray}
where each $\bar{M}_{i}^{3}(t)$ and $\bar{M}_{i}^{4}(t)$ are the two free functions that contribute at $i$-th order in the action. This is a logical extension to $n$-th order of 
the first two operators which appear in $\mathcal{S}^{(2)}$ in Eq.~\eqref{eq:s2}, namely $\small(\delta g^{00}\small)^{2}$ and $\delta g^{00}\delta K$.

%%%%%% SCREENING %%%%%%%
\section{Nonlinear freedom for screening} \label{sec:screening}
%%%%%%%%%%%%%%%%%%%%%%%%

%
As a first application of the free nonlinear correction term in Eq.~\eqref{DeltaG23} in the reconstructed scalar-tensor action we shall consider the realization of screening mechanisms that are required to recover GR in the well-tested Solar-System regime \cite{Will:2014kxa}.
For this purpose, we shall employ the scaling method of Refs.~\cite{McManus:2016kxu, McManus:2017itv}
(also see applications in Refs.~\cite{Lombriser:2016zfz,2018IJMPD..2748002L,Bolis:2018kcq})
that allows an efficient identification of the existence of Einstein gravity regimes for a particular choice of Horndeski functions.
We briefly review the method (Sec.~\ref{sec:scalingmethod}) and then apply it for a characterization of the nonlinear correction terms $\Delta G_i$ that realize screening by large gravitational potentials $\Phi_N>\Lambda$ for some threshold $\Lambda$ (Sec.~\ref{sec:chameleon}), large first derivatives $\nabla\Phi_N>\Lambda$ (Sec.~\ref{sec:kmouflage}) or large second derivatives $\nabla^2\Phi_N>\Lambda$ (Sec.~\ref{sec:vainshtein})~\cite{Joyce:2016vqv}.

%------ SCALING -------%
\subsection{Scaling method} \label{sec:scalingmethod}
%----------------------%

The scaling method was developed in Refs.~\cite{McManus:2016kxu, McManus:2017itv} to efficiently determine whether a given Horndeski theory possesses an Einstein gravity limit.
It proceeds as follows.
At the level of the field equations the scalar field $\phi$ is expanded in terms of a field perturbation $\psi$ as
\begin{equation}
    \phi=\phi_{0}\left(1+\alpha^{q}\psi \right) \, ,
    \label{scalingrelation}
\end{equation}
where $\phi_{0}$ denotes the background value and $\alpha$ is the theoretical parameter relevant to the expansion. For example, it could be the speed of light or the coupling of a Galileon interaction term.
After performing this expansion, the scalar field equation of the Horndeski model (see Eq.~\eqref{SFequation}) takes the generic form 
\begin{equation}
    \alpha^{s+mq}F_{1}(\psi,\tilde{X})+\alpha^{t+nq}F_{2}(\psi,\tilde{X})=\frac{T}{M_{*}^{2}} \,, 
    \label{HorndeskiFEafterscaling}
\end{equation}
where $s,m,t,n\in \mathbb{N}$ and $\tilde{X}=\partial_{\mu}\psi\partial^{\mu}\psi$.
Now consider the limit of $\alpha \rightarrow \infty$ or $\alpha \rightarrow 0$.
As the right-hand side of Eq.~\eqref{HorndeskiFEafterscaling} is independent of $\alpha$ the leading-order term on the left-hand side must also be independent of $\alpha$ to balance the equation.
This restricts the possible values of the exponent $q$.
Therefore there must be at least one term which scales as $\alpha^{0}$ with every other term involving non-zero powers of $\alpha$ vanishing in the $\alpha \rightarrow 0$ or $\alpha \rightarrow \infty$ limit. For example, choosing $q=-s/m$ and taking the $\alpha \rightarrow \infty$ limit requires $-s/m<-t/q$, so that $t+n(-s/m)<0$ and the dominating terms in the field equation become
\begin{equation}
    F_{1}(\psi,\tilde{X})=\frac{T}{M_{*}^{2}} \, .
    \label{SFquationsreenedlimit}
\end{equation}
If in a given $\alpha$ limit the metric field equations reduce to the Einstein equations after performing the expansion~\eqref{scalingrelation}, then the corresponding scalar field equation applies to the screened limit where the fifth force is suppressed.
To ensure consistency the value of $q$ chosen to obtain a screened limit must be the same in both the scalar and metric field equations.
Note that there may also be terms that involve powers of $\alpha$ that do not depend on $q$. Depending on whether they are raised to a positive or a negative power they will diverge or vanish in either limit of $\alpha$. If they vanish then this is not an issue, but if they diverge extra care must be taken. For example, it may be important to use the freedom in the $\Delta G_{i}$ terms to remove any divergences which arise in either limit.

In the following we present the recovery of three distinct screening mechanisms by suitable choices of $\Delta G_i$.
Drawing on the distinction discussed in Ref.~\cite{Joyce:2016vqv} this will encompass the known screening mechanisms: (i) by large gravitational potentials $\Phi_N>\Lambda$ for some threshold $\Lambda$ (Sec.~\ref{sec:chameleon}), (ii) by large first derivatives $\nabla\Phi_N>\Lambda$ (Sec.~\ref{sec:kmouflage}) and (iii) by large second derivatives $\nabla^2\Phi_N>\Lambda$ (Sec.~\ref{sec:vainshtein}).
We shall find that there is more than sufficient freedom in the nonlinear sector to, in principle,
endow the reconstructed theory with a particular screening mechanism regardless of the constraints of the background and the linear perturbations.
Importantly, however, while this generally implies the existence of Einstein gravity limits in the deeply nonlinear regime, this does not guarantee that a given observed region is nonlinear enough for the screening mechanism to be activated.
The numerical value of the screening scale needs to be computed separately and ultimately decides whether a theory is compatible with stringent Solar-System tests.
It is not surprising that screening mechanisms can be added to linearly reconstructed models as they are inherently nonlinear effects.
It is however important to verify
this explicitly. 

%----- CHAMELEON ------%
\subsection{Large field value screening} \label{sec:chameleon}
%----------------------%

As a first example we consider the implementation of a screening effect by large field values $\Phi_N>\Lambda$.
More specifically, we will focus on the Chameleon Mechanism \cite{Khoury:2003aq, PhysRevD.69.044026}.
We shall first cast the reconstructed theory
into the Brans-Dicke representation with $F(\phi)=\phi/M_{*}$ (see Sec.~III in Ref.~\cite{Kennedy:2018gtx}).
With this choice we have that $\Gamma=\phi/M_{*}$ and $\Xi=1$ in Eqs.~\eqref{MFequation} and \eqref{SFequation}.
By making use of the freedom in $\Delta G_{i}$ it is possible to add a term to $G_{2}$ that sets the $q$-value to be arbitrarily positive or negative.
To see this let us begin with the full reconstructed Horndeski action in Eqs.~\eqref{eq:G2recon} to \eqref{eq:G4recon} with a $\Delta G_{2}$ term that takes the form
\begin{equation}
    \Delta G_{2}=\xi(\phi)\left(1+\frac{X}{M_{*}^{4}}  \right)^{n} \, ,
    \label{Chameleoncounterterm}
\end{equation}
with $n\geq 3$ and $\xi(\phi)$ given by
\begin{equation}
    \xi(\phi)=M_{*}^{2}U(\phi)-\frac{\lambda^{-N}}{2}\left(\phi-\phi_{min} \right)^{k} \,,
    \label{xichoiceChameleon}
\end{equation}
where $\lambda$ is a coupling parameter, $N$ and $k$ are both positive integers, $U(\phi)$ is the reconstructed potential in Eq.~\eqref{eq:G2recon} and $\phi_{min}$ denotes the minimum value of the second contribution to the potential in Eq.~\eqref{xichoiceChameleon}. No other $\Delta G_{i}$ terms are necessary as they all contain derivative terms which vanish in the screened limit.
We shall take the scaling parameter $\alpha$ to be the coupling $\lambda$.

This choice cancels the potential obtained from the linear reconstruction and replaces it with a power-law potential that takes a similar form to the chameleon screening example in Ref.~\cite{McManus:2016kxu,Lombriser:2013eza} but with $\alpha\rightarrow\alpha^{-N}$.
It is with a suitable choice of $N$ that no derivative terms contribute in the screening limit.
In this limit we then obtain the Einstein equation
\begin{equation}
    \frac{\phi}{M_{*}}R_{\mu\nu}=-\mathcal{T}^{(2)}_{\mu\nu}+\left(T_{\mu\nu}-\frac{1}{2}g_{\mu\nu}T \right)/M_{*}^{2} +H_{m}\left[\nabla_{\mu}\phi\right]\, ,
    \label{Chammetricfieldequation}
\end{equation}
where $\mathcal{T}^{(2)}_{\mu\nu}$ is defined in Eq.~\eqref{Eq6} and $H_{m}\left[\nabla_{\mu}\phi\right]$ represents all the terms that involve derivatives of $\phi$ in the metric field equation, the precise form of which is not relevant as we shall find that they disappear in the $\alpha \rightarrow 0$ limit of interest.
Taking the trace of Eq.~\eqref{Chammetricfieldequation} leads to $\phi R/M_{*}=-\mathcal{T}^{(2)}$ which, noting that $\mathcal{T}^{(2)}=2G_{2}/M_{*}^{2}$, gives a relation between $R$ and $G_{2}$. The scalar field equation is given by
\begin{equation}
    -\frac{2\phi}{M_{*}^{2}}\left(G_{2\phi}+G_{4\phi}R \right)+\mathcal{T}^{(2)}+H_{s}\left[\nabla_{\mu}\phi\right]=-T/M_{*}^{2} \, ,
\end{equation}
where $H_{s}\left[\nabla_{\mu}\phi\right]$ represents all the terms in the scalar field equation involving derivatives of $\phi$ which will disappear in the $\alpha \rightarrow 0$ limit.
With the choice of $\Delta G_{2}$ in Eq.~\eqref{Chameleoncounterterm} there is no contribution from the reconstructed potential $U(\phi)$ to the scalar field equation. After eliminating $R$ and $\mathcal{T}^{(2)}$ in favour of $G_{2}$ the scalar field equation becomes
\begin{equation}
    \alpha^{-N}\left(\phi-\phi_{min} \right)^{k-1}\left[  \phi k -2\left(\phi-\phi_{min}\right)\right]+H_{s}\left[\nabla_{\mu}\phi\right]=-T \, .
\label{SFequation2}
\end{equation}
Applying the scaling method with the scalar field now expanded in terms of $\psi$ as in Eq.~\eqref{scalingrelation}, we examine the set of $q$ values which leave non-vanishing terms on the left-hand side of Eq.~\eqref{SFequation2} in the $\alpha \rightarrow 0$ limit.
As $\alpha \rightarrow 0$ it is necessary to take the largest $q$ value from this set after the scaling in Eq.~\eqref{SFequation2}.
Disregarding the derivative terms in $H_{s}\left[\nabla_{\mu}\phi\right]$, we find that $q$ takes one of two possible values
\begin{equation}
    q \in \left\{\frac{N}{k-1},\frac{N}{k} \right\} \, .
\end{equation}
We must take $q=N/(k-1)$ as it is the largest in the set of $q$ values from $G_{2}$. The integer $N$ can then be chosen in Eq.~\eqref{xichoiceChameleon} to be arbitrarily large.
In the limit of $\alpha\rightarrow 0$ this will send all terms involving spacetime derivatives of $\phi$ to zero, justifying the original choice of $\xi(\phi)$. This is important as in principle the value of $n$ in Eq.~\eqref{Chameleoncounterterm} is only bounded from below by the requirement that it is a nonlinear correction.
All the terms involving derivatives of the scalar field scale as $X^{m}=\phi_{0}^{2m}\alpha^{2mN/(k-1)}\tilde{X}\rightarrow 0$ as $\alpha \rightarrow 0$ with $m=\left\{1, \, \ldots \, ,n   \right\}$.

Now we expand the scalar field around the minimum of the potential such that $\phi_{min}\approx \phi_{0}$. This then implies that $\phi-\phi_{0}=\phi_{0}\alpha^{q}\psi$.
The remaining terms
in the scalar field equation
for $\alpha\rightarrow 0$
relate the local value of the scalar field to the matter density as
\begin{equation}
    \psi=\left(\frac{-T}{\phi_{0}^{k}k} \right)^{\frac{1}{k-1}} \,,
\end{equation}
which recovers the chameleon screening effect for $k<1$.
The metric field equation in the same limit reduces to %
\begin{equation}
    \phi_{0}R_{\mu\nu}=\left(T_{\mu\nu}-\frac{1}{2}g_{\mu\nu}T\right)/M_{*} \, ,
    \label{Einsteinequation}
\end{equation}
recovering the standard Einstein equation with a re-scaled Planck mass
set by the background field value $\phi_{0}$.
Therefore we have implemented a Chameleon Mechanism in a scalar-tensor action that is reconstructed from an arbitrary cosmological background evolution and linear perturbations by adding a suitable choice of $\Delta G_{2}$.
Whether the screening effect operates in the Solar System to comply with stringent local tests of gravity needs to be checked numerically for a given reconstructed model.
%

%----- K-MOUFLAGE -----%
\subsection{First-derivative screening} \label{sec:kmouflage}
%----------------------%

Next we examine the implementation of a screening effect that operates through large first derivatives $\nabla\Phi_N>\Lambda$.
More specifically, we focus on the k-mouflage screening effect~\cite{Babichev:2009ee, Brax:2012jr}.
We may simply choose here the scaling parameter $\alpha$ to be the kineticity function $\alpha_{K}$ and take the $\alpha\rightarrow \infty$ limit.
EFT functions such as $\alpha_{K}$ are typically parameterised as $\alpha_{K0}f(a)$ where $f(a)$ is some function of the scale factor with $f(a=1)\equiv1$. Often this is simply a power of the scale factor or the evolution of the dark energy density normalised to the present value $\Omega_{DE}(a)/\Omega_{DE0}$. This ensures that the effects of the modifications only become relevant at late times.
We shall take here the scaling parameter to correspond to
the value of $\alpha_{K}$ today $\alpha=\alpha_{K0}$. It is also possible to take $\alpha_{B0}$ or $\alpha_{M0}$ as the scaling parameter but as the reconstruction depends differently on these EFT parameters this will lead to different behaviour in the screened limit (see Sec.~\ref{sec:vainshtein}).
Taking $\alpha$ to be $\alpha_{K0}$, we see that as the reconstructed action is linear in the EFT functions we have from Table~\ref{tab:solution} that each term scales as $U(\phi)\sim \alpha$, $Z(\phi)\sim \alpha$, $a_{2}(\phi)\sim \alpha$ and $b_{1}(\phi)\sim \alpha^{0}$, which follows from the fact that $\Bar{M}_{1}^{3}$ is independent of $\alpha_{K}$ (see Table~II in Ref.~\cite{Kennedy:2017sof} for the full set of relations between the EFT coefficients of the different bases). 
With this choice we have that the terms in $G_{2}$ will scale as $\alpha^{1+nq}$ for some integer $n$ but those in $G_{3}$ will scale as $\alpha^{nq}$.

In order to obtain an Einstein field equation it is necessary to remove the potential to avoid divergences in the $\alpha \rightarrow \infty$ limit. This also makes physical sense as the screening mechanism in this case operates via the kinetic terms.
We shall also remove all of the dependence on the canonical kinetic term linear in $X$ to ensure that the screening operates through higher powers of $X$.
To this end, we choose $\Delta G_{2}=\Delta G^{(1)}_{2}+\Delta G^{(2)}_{2}$, where 
\begin{equation}
    \Delta G^{(1)}_{2}=\frac{1}{2}M_{*}^{6}Z(\phi)\left(1+\frac{X}{M_{*}^{4}}   \right)^{4}- \frac{1}{2}M_{*}^{6}Z(\phi)\left(1+\frac{X}{M_{*}^{4}}   \right)^{3} \,,
    \label{correction1}
\end{equation}
\begin{equation}
    \Delta G_{2}^{(2)}=2M_{*}^{2}U(\phi)\left(1+\frac{X}{M_{*}^{4}}   \right)^{3}- M_{*}^{2}U(\phi)\left(1+\frac{X}{M_{*}^{4}}   \right)^{6} \,.
    \label{correction2}
\end{equation}
These nonlinear corrections ensure that every term in $G_{2}$ is now at least proportional to $X^{2}$ or greater.
With this choice the relevant term in the scalar field equation is 
\begin{equation}
    \nabla^{\mu}J^{(2)}_{\mu}=-G_{2XX}\nabla^{\mu}X\nabla_{\mu}\phi -X G_{2X\phi} \, ,
    \label{kmouflagescalingexample}
\end{equation}
where $J^{(2)}_{\mu}$ is defined in Eq.~\eqref{Eq4}. The first term on the right-hand side in Eq.~\eqref{kmouflagescalingexample} scales as $\alpha^{1+3q}$, which sets the minimum $q$-value to be $q=-1/3$. As every term in $G_{3}$ scales as $\alpha^{nq}$ with $n>0$ this will send every term involving $G_{3}$ to zero in the $\alpha \rightarrow \infty$ limit. This particular $q$-value will also ensure that $\mathcal{T}^{(i)}_{\mu\nu} \rightarrow 0$ as $\alpha \rightarrow \infty$ so that the metric field equation reduces to the standard Einstein field equation. The resulting scalar field equation corresponds to a k-mouflage model
\begin{equation}
   \xi(\phi)\partial^{\mu}X\partial_{\mu}\phi=-\frac{T}{M_{*}^{2}} \, ,
\end{equation}
with 
\begin{equation}
    \xi(\phi)= a_{2}(\phi)+\frac{9Z(\phi)}{2M_{*}^{2}}-\frac{9U(\phi)}{M_{*}^{8}} \, .
\end{equation}
%

%----- VAINSHTEIN -----%
\subsection{Second-derivative screening} \label{sec:vainshtein}
%----------------------%

%
Finally, we consider the implementation of screening through large second derivatives $\nabla^2\Phi_N>\Lambda$, more explicitly the realization of the Vainshtein mechanism in the $\alpha \rightarrow \infty$ limit
where the scaling parameter $\alpha$ is taken to be $\alpha_{B}$ only.
The procedure is similar to Sec.~\ref{sec:kmouflage}. In this case $U(\phi)\sim \alpha$, $Z(\phi)\sim \alpha$, $a_{2}(\phi)\sim \alpha$ as before, but in contrast to Sec.~\ref{sec:kmouflage}, $b_{1}(\phi)\sim \alpha$, which follows from the fact that $\Bar{M}_{1}^{3} \propto \alpha_{B}$. 
We begin by adding on the nonlinear counterterms in Eqs.~\eqref{correction1} and \eqref{correction2} to ensure the $X$ dependence of $G_{2}$ is at least $X^{2}$. 

It turns out that  the important term in the scalar field equation which gives rise to a non-trivial equation of motion and Vainshtein screening is $\nabla^{\mu}J^{(3)}_{\mu}$ where $J^{(3)}_{\mu}$ is given in Eq.~\eqref{Eq5}. Plugging in the expression in Eq.~\eqref{eq:G3recon} we have that
\begin{equation}
    \nabla^{\mu}J^{(3)}_{\mu}=b_{1}(\phi)\mathcal{S}^{(4,2)}+H_{s}\left[\nabla_{\mu}\phi\right] \, ,
\end{equation}
where again $H_{s}\left[\nabla_{\mu}\phi\right]$ represents all of the terms involving derivatives of $\phi$ that will vanish in the $\alpha \rightarrow \infty$ limit. Furthermore $S^{(4,2)}$ is a term that involves four derivative operators and two powers of the scalar field, which is given explicitly by
\begin{equation}
    \mathcal{S}^{(4,2)}=\left(\Box \phi \right)^{2}+\partial_{\mu}\phi
\partial^{\mu}\Box \phi+\Box X \, .
\end{equation}
These terms each scale as $\alpha^{1+2q}$ requiring a $q$-value of $-1/2$ to ensure independence of $\alpha$ on the left-hand side. As we have also ensured that $G_{2}$ starts at least at $X^{2}$, scaling as $\alpha^{4q}$ with $q=-1/2$, these higher-derivative terms will disappear in the $\alpha \rightarrow \infty$ limit. The scalar field equation in this limit then becomes
\begin{equation}
    \frac{\phi_{0}^{3}}{M_{*}}b_{1}(\phi_{0})\left[ \left(\Box \psi \right)^{2}+\partial_{\mu}\psi
\partial^{\mu}\Box \psi+\Box \tilde{X}  \right] =-\frac{T}{M_{*}^{2}}  \, ,
\end{equation}
where $\tilde{X}\equiv \partial_{\mu}\psi\partial^{\mu}\psi $. This is a typical scalar field equation involving higher derivatives of $\psi$ expected for Vainshtein screening. It is necessary to ensure that the standard Einstein equation is obtained in the same limit in the metric field equations so that we can be sure this is the screened limit. 

Having already set $q=-1/2$ from the scalar field equation and ensured that $G_{2}$ starts at $X^{2}$ with $\Delta G_{2}^{(1)}$ and $\Delta G_{2}^{(2)}$, every term $\mathcal{T}^{(i)}_{\mu\nu}$ in the metric field equation \eqref{MFequation} vanishes in the $\alpha \rightarrow \infty$ limit. For example the first term in $\mathcal{T}^{(4)}_{\mu\nu}$ scales as
\begin{equation}
    G_{4\phi}\mathcal{S}^{(2,1)} \sim \alpha^{2q} \sim \alpha^{-1} \rightarrow 0 \,,
\end{equation}
and the first one in $\mathcal{T}^{(3)}_{\mu\nu}$ scales as
\begin{equation}
      \frac{2}{M_{*}^{2}}G_{3X}\mathcal{S}^{(4,3)} \sim \alpha^{1+3q} \sim \alpha^{-\frac{1}{2}} \rightarrow 0 \, .
\end{equation}
With the choice of the Brans-Dicke representation of $F(\phi)=\phi/M_{*}$ we have that $\Gamma = \phi_{0}/M_{*}$ and $\Xi=1$, and the metric field equation reduces to Eq.~\eqref{Einsteinequation}.

To summarize, by choosing $\alpha_{B0}$ as the scaling parameter and removing the constant and linear terms in $X$ from $G_{2}$ one can obtain the standard Einstein field equation with a re-scaled Planck mass and a scalar field equation involving second derivatives in $\psi$ as expected in the case of Vainshtein screening. 

%%%%%% KINETIC-SA %%%%%%
\section{Nonlinear freedom for degenerate kinetic self-acceleration} \label{sec:kineticSA}
%%%%%%%%%%%%%%%%%%%%%%%%

As a further application of the nonlinear freedom in reconstructed scalar-tensor theories, we demonstrate how the correction term in Eq.~\eqref{DeltaG23} can be configured to construct scalar-tensor theories that are degenerate with standard cosmology to an arbitrary level of cosmological perturbations (Sec.~\ref{sec:degeneracy}).
As a particular interesting example we show how this allows for models that accelerate the Universe without a cosmological constant yet remain dynamically degenerate with $\Lambda$CDM through a suitable configuration of the kinetic terms (Sec.~\ref{sec:kinselfacc}).

%----- DEGENERACY -----%
\subsection{Perturbative degeneracy with $\Lambda$CDM} \label{sec:degeneracy}
%----------------------%

An important implication of Eq.~\eqref{DeltaG23} is that it is possible to use the $\Delta G_{i}$ terms to write down a Horndeski theory that possesses a highly non-trivial form for the nonlinear perturbations yet reduces to $\Lambda$CDM on the background, where the correction terms vanish.
This degeneracy may even be extended to an arbitrary level of perturbations.
The existence of such classes of theories is a natural consequence of the reconstruction being an expansion in $\left(1+X/M_{*}^{4} \right)^{n}$ with $n\in \mathbb{N}$. One can therefore construct theories whose physical effects only become relevant at a particular level of higher-order perturbations characterized by the power $n$.

To see how this works in practice let us choose, for example,
\begin{equation}
    G_{2}=-M_{*}^{2}\Lambda+\xi^{{\scriptscriptstyle(2)}}_{n}(\phi)\left(1+\frac{X}{M_{*}^{4}}\right)^{n} \, ,
    \label{kinSAexample}
\end{equation}
with $G_{3}=0$, $G_{4}=M_{*}^{2}/2$ and $n\geq 3$.
After performing an ADM decomposition with $\phi=t M_{*}^{2}$ the second term in Eq.~\eqref{kinSAexample} becomes $\xi^{{\scriptscriptstyle(2)}}_{n}(t)\left(\delta g^{00} \right)^{n}$. On the background and linear scales therefore there will be no effects arising from the non-canonical kinetic terms and it will appear to be exactly $\Lambda$CDM.
Note that this argument does not rely on the specific foliation adopted as we shall verify shortly for a specific example, but for now simply note that any non-zero perturbations that arise from another choice of foliation must be pure gauge.
At the nonlinear level Eq.~\eqref{kinSAexample} departs from $\Lambda$CDM and we have discussed the mapping of the $\xi^{{\scriptscriptstyle(2)}}_{n}(t)$ functions onto nonlinear EFT functions in Sec.~\ref{sec:higherorderreconstruction}. It is also possible to write a theory with $G_{2}=\Lambda$ and 
\begin{equation}
    G_{3}=\xi^{{\scriptscriptstyle(3)}}_{n}(\phi)\left(1+\frac{X}{M_{*}^{4}}\right)^{n} \, .
\end{equation}
In an equivalent manner this corresponds to a Galileon theory that can only be distinguished from $\Lambda$CDM on nonlinear scales. Combinations of $\Delta G_{2}$ and $\Delta G_{3}$ can also be used to construct more non-trivial theories. 

For clarity we shall provide an explicit example of this degeneracy and compute the background equations of motion and check that the expansion is indeed matching that of $\Lambda$CDM. A more detailed analysis, including the investigation of possible instabilities and perturbative effects, will be the subject of further analysis. 
For simplicity, we shall only focus here on the degeneracy at the level of the background and not for the linear perturbations.
Hence, we take $n=2$ in Eq.~\eqref{kinSAexample} so that
\begin{align}
    G_{2}&=-M_{*}^{2}\Lambda+\xi(\phi)\left(1+\frac{X}{M_{*}^{4}} \right)^{2} \, , \nonumber\\
    &=-M_{*}^{2}\Lambda+\xi(\phi)+2\xi(\phi)X/M_{*}^{4}+\xi(\phi)X^{2}/M_{*}^{8} \, ,
    \label{DegenerateKSA}
\end{align}
where $\xi(\phi)$ is a free function of $\phi$.
Not making any assumptions about the space-like foliation we now put this equation into the unitary gauge by setting the scalar field to be just a function of time. With $X=\left(-1+\delta g^{00} \right)\dot{\phi}^{2}$ we have at linear order
\begin{align}
    G_{2}=&-M_{*}^{2}\Lambda+\xi(t)+\frac{2\xi(t)X_{0}}{M_{*}^{4}}+\frac{\xi(t)X_{0}^{2}}{M_{*}^{8}} \nonumber \\
    &-\left[\frac{2\xi(t)X_{0}}{M_{*}^{4}} + \frac{2\xi(t)X_{0}^{2}}{M_{*}^{8}} \right]\delta g^{00}                 \, ,
    \label{expansionofG2}
\end{align}
where $X_{0}$ is the value that $X$ takes on the background, i.e., $X_{0}=-\dot{\phi}^{2}$. This gives an explicit expression for the EFT functions $\Lambda(t)$ and $\Gamma(t)$ in the unitary gauge expansion of $G_{2}$ in Eq.~\eqref{expansionofG2}, where first line corresponds to $-M_{*}^{2}\Lambda(t)$ and the second line to $-M_{*}^{2}\Gamma(t)/2$. 
%
%These are given by
%
%\begin{eqnarray}
% &\Lambda(t)& = \Lambda-\frac{\xi(t)}{M_{*}^{2}}-\frac{2\xi(t)X_{0}}{M_{*}^{6}}-\frac{\xi(t)X_{0}^{2}}{M_{*}^{10}} \, , \label{Lambdaoft} \\
% &\Gamma(t)& = \frac{4\xi(t)X_{0}}{M_{*}^{6}}+\frac{4\xi(t)X_{0}^{2}}{M_{*}^{10}} \, . 
% \label{Gammaoft}
%\end{eqnarray}
%
The Friedmann equations in the EFT formulation are given by~\cite{Gubitosi2012, Lombriser:2014ira, Gleyzes:2014rba}
\begin{eqnarray}
&\Gamma(a) &+\Lambda(a)  = 3H^{2}-\frac{\rho_{m}}{M_{*}^{2}} \,, \label{FE1} \\
& \Lambda(a) & = 2HH^{\prime}+3 H^{2}  \label{FE2}\,,
\end{eqnarray}
where we have set the non-minimal coupling parameter $\Omega=1$, we parameterise the time $t$ in terms of the scale factor $a$. With the expressions for $\Gamma(a)$ and $\Lambda(a)$ obtained from Eq.~\eqref{expansionofG2} one can take linear combinations of the Friedmann equations~\eqref{FE1} and \eqref{FE2} to eliminate the dependence on the background expansion $H$ and obtain a field equation for the background value of the scalar field. This is determined from the resulting expression
\begin{equation}
    \Gamma(a)+\frac{1}{3}\left[\Gamma(a)+\Lambda(a) \right]^{\prime}=0
\end{equation}
to be
\begin{align}
&\left[\frac{4\xi(a)X_{0}}{M_{*}^{10}}+\frac{\xi^{\prime}(a)}{M_{*}^{10}}\left(X_{0}-M_{*}^{4}/3 \right) \right]\left(X_{0}+M_{*}^{4} \right) \\ 
&+X_{0}^{\prime}\left[\frac{2\xi(a)}{3M_{*}^{6}}+\frac{2\xi(a)X_{0}}{M_{*}^{10}}   \right]=0 \, ,
\end{align}
which is the non-trivial Klein-Gordan scalar field equation. It has a trivial solution $X_{0}=-M_{*}^{4}$.
More complicated solutions to the background scalar field equation will be explored in the future.
From $X_{0}=-M_{*}^{4}$, one immediately recognizes in Eq.~\eqref{DegenerateKSA} that $G_2(X_0) = -\Lambda$, and hence the recovery of the $\Lambda$CDM background expansion.
Alternatively, once the solution to the background evolution of the scalar field has been obtained it is possible to derive the equation-of-state parameter for the resulting k-essence model given by \cite{Cordero:2016bxt}
\begin{equation}
    w(a)=\frac{-M_{*}^{2}\Lambda+\xi(\phi)\left(1+X/M_{*}^{4} \right)^{2}}{M_{*}^{2}\Lambda-\xi(\phi)\left(1+X/M_{*}^{4} \right)\left(1-3X/M_{*}^{4} \right)} \, .
\end{equation}
After inserting the background solution $X=X_{0}=-M_{*}^{4}$ one obtains $w=-1$, confirming that the background expansion is indeed matching that of $\Lambda$CDM.

%----- KinSelfAcc -----%
\subsection{Degenerate kinetic self-acceleration} \label{sec:kinselfacc}
%----------------------%

%
To highlight the implications of the perturbative degeneracy, we will now study a particularly interesting example of Eq.~\eqref{kinSAexample}. Let us consider a class of models specified by $\xi(\phi)=M_{*}^{2}\Lambda_{\phi}$ in Eq.~\eqref{DegenerateKSA}. The subscript $\phi$ indicates that $\Lambda_{\phi}$ is a coupling parameter in the higher-order kinetic terms of the scalar field $\phi$. Eq.~\eqref{DegenerateKSA} then becomes
\begin{equation}
  G_{2}  =-M_{*}^{2}\Lambda_{GR}+M_{*}^{2}\Lambda_{\phi}\left(1+ \frac{X}{M_{*}^{4}}   \right)^{2} \,, %\, n\geq 3 \, ,
  \label{eq:kineticselfacc}
\end{equation}
where we defined $\Lambda\equiv\Lambda_{GR}$.
We also set $G_{4}=1$ and $G_{3}=0$ and stress that any contributions to $\Lambda_{GR}$ from quantum corrections of matter fields in this discussion are neglected. 
If we now set $\Lambda_{\phi}=\Lambda_{GR}$ this model exhibits the particular feature of having no explicit cosmological constant. The model is now simply
\begin{equation}
    G_{2}=2\Lambda_{\phi}X/M_{*}^{2}+\Lambda_{\phi}X^{2}/M_{*}^{6} \,.
\end{equation}
However, the observed cosmological constant $\Lambda_{obs}$ in the cosmological background of this model remains $\Lambda_{obs}=\Lambda_{\phi}=\Lambda_{GR}$.
An alternative approach is to start with the model
\begin{equation}
    G_{2}=2\Lambda_{\phi}X/M_{*}^{2}+\Lambda_{\phi}X^{2}/M_{*}^{6} -2M_{*}^{2}\Lambda_{GR}\, ,
\end{equation}
and then set $\Lambda_{GR}=0$. In summary, in one interpretation the coupling $\Lambda_{\phi}$ is tuned to match a non-vanishing $\Lambda_{GR}$ that corresponds to the observed $\Lambda_{obs}$ or $\Lambda_{GR}=0$ and $\Lambda_{\phi}=\Lambda_{obs}$.

With either interpretation these models generate a \emph{kinetic self-acceleration} effect that is degenerate with the cosmological constant to the $(n-1)$-th order of cosmological perturbations.
While this may certainly be viewed as an engineered self-acceleration effect, it also raises more general questions about the genuineness of a kinetic self-acceleration that resembles a cosmological constant for observational compatibility.
We note that a similar expansion to Eq.~\eqref{eq:kineticselfacc} can be performed for $G_3$ with similar implications.
For instance, one may consider a kinetic gravity braiding model with nontrivial $G_2$ and $G_3$.
By combining power series of $(1+X/M_{*}^{4})^{n}$ in $G_2$ and $G_3$ that only contribute at $(n-1)$-th order in cosmological perturbations, one can choose the coefficients of $G_2$ and $G_3$ in an expansion in $X$ to cancel off to just leave a term $X^{n}$ in $G_{2}$ and $G_{3}$ for arbitrarily large $n$.
Greater values of $n$ then correspond to models which are more difficult to distinguish from $\Lambda$CDM and for which nonlinear data must be used for their discrimination.
This may shed some light on the results of Ref.~\cite{Kimura:2011td}, where better agreement with $\Lambda$CDM at the linear level was likewise found for kinetic gravity braiding models with $G_3\propto X^n$ for large $n$ but adopting a canonical $G_2$ instead, which is not feasible with using $\Delta G_{i}$ corrections only.

We shall leave a more detailed examination of the genuineness of kinetic self-acceleration that closely matches $\Lambda$CDM phenomenology to subsequent work.
It is worth noting however that a further interesting consequence of $\Lambda_{obs}$ being interpreted as a coupling rather than a bare constant is that it may be possible to render the acceleration effect in Eq.~\eqref{eq:kineticselfacc} technically natural as it can now enter as a coefficient to an irrelevant operator rather than as a non-renormalizable constant~\cite{Gripaios:2015qya, Pirtskhalava:2015nla}. The details shall also be studied further in forthcoming work.
At a more practical level, we emphasise that these models have the interesting property that discriminatory effects of this type of cosmic acceleration are left exclusively to the nonlinear observational regime.

%-- HIGHER-ORDER RECO -%
\section{Higher-order reconstruction} \label{sec:higherorderreconstruction}
%----------------------%

With the higher-order EFT expansion in Eq.~\eqref{nthorderEFT} and the freedom in the nonlinear sector having been significantly reduced by the restriction to a luminal speed of gravity, it becomes straightforward to perform a $n$-th order reconstruction of the corresponding class of Horndeski theories by fixing the $\Delta G_{i}$ functions order-by-order in terms of the nonlinear EFT functions $\bar{M}_{i}^{3,4}$.
We shall now see how this extra information modifies the reconstruction from the background and linear scales by adding in the new free functions and slightly changing the dependence on the linear EFT functions.
We shall elaborate on this explicitly for the case of $i=3$ before outlining the general $n$-th order case. 

Let us begin by noting that in the unitary gauge a term that takes the form $\xi(\phi)X^{m}\Box \phi$ becomes
\begin{align}
\xi(\phi)X^{m}\Box \phi=&\mp \frac{2m}{2m+1}\xi(\phi)(-X)^{m+\frac{1}{2}}K \nonumber \\ &\pm\frac{1}{2m+1}\xi^{\prime}(\phi)(-X)^{m+1} \,, 
\label{XmBoxphi}
\end{align}
where the sign difference on the top and bottom indicate even or odd $m$ respectively and the prime denotes a derivative with respect to $\phi$.
After expanding Eq.~\eqref{XmBoxphi} in the unitary gauge there will be several terms that contribute and that can be mapped onto the operators in Eq.~\eqref{nthorderEFT}.

We shall proceed along the same lines as Ref.~\cite{Kennedy:2017sof} to obtain a corresponding covariant action. To begin, by using the replacement $\delta g^{00}=1+X/M_{*}^{2}$ the $\left(\delta g^{00}\right)^{3}$ operator becomes 
\begin{equation}
    \bar{M}_{3}^{4}(t)\left(\delta g^{00}\right)^{3}= \bar{M}_{3}^{4}(\phi)\left(1+\frac{3X}{M_{*}^{4}}+\frac{3X^{2}}{M_{*}^{8}}+\frac{X^3}{M_{*}^{12}} \right) \, .
\end{equation}
This contributes to $U(\phi)$, $Z(\phi)$, $a_{2}(\phi)$ along with a new, now necessarily non-vanishing contribution to the coefficient of $X^{3}$ that we call $a_{3}(\phi)$.
Let us now derive the covariant action which gives rise to the following expansion in the unitary gauge
\begin{equation}
    \bar{M}_{1}^{3}(t)\delta g^{00}\delta K+\bar{M}_{3}^{3}(t)(\delta g^{00})^{2}\delta K \, .
\end{equation}
We shall take the case of $m=1,2$ in Eq.~\eqref{XmBoxphi} for simplicity and begin with the combination
\begin{equation}
  G_{3} = b_{1}(\phi)X \Box \phi+b_{2}(\phi)X^{2}\Box\phi+\Delta G_{3}^{(4)} \, , 
  \label{UGexpansionb1b2}
\end{equation}
where $\Delta G_{i}^{(4)}$ indicates that the nonlinear corrections now start at fourth order.
We transform Eq.~\eqref{UGexpansionb1b2} into the unitary gauge and then solve for $b_{1}(\phi)$ and $b_{2}(\phi)$ in terms of the EFT functions. It is necessary to have two independent functions in the covariant expansion as there are two independent EFT functions. At third order in the perturbations we obtain
\begin{align}
    G_{3}\supset &-b_{1}(\phi)M_{*}^{6}\delta g^{00}\delta K+\frac{1}{4}b_{1}(\phi)M_{*}^{6}(\delta g^{00})^{2}\delta K\\ &+2b_{2}(\phi)M_{*}^{10}\delta g^{00}\delta K-\frac{3}{2}b_{2}(\phi)M_{*}^{10}(\delta g^{00})^{2}\delta K \, ,
\end{align}
where for the sake of clarity we have not shown the terms which are independent of $\delta K$. We then require that 
\begin{equation}
    -b_{1}(\phi)M_{*}^{6}+2b_{2}(\phi)M_{*}^{10}=\bar{M}_{1}^{3}(\phi) \, ,
\end{equation}
\begin{equation}
    b_{1}(\phi)M_{*}^{6}-6b_{2}(\phi)M_{*}^{10}=4\bar{M}_{3}^{3}(\phi) \, .
\end{equation}
This system of equations can be straightforwardly solved to obtain $b_{1}(\phi)$ and $b_{2}(\phi)$.
The results are shown in Table~\ref{nonlinearsolution} along with the contributions to $G_{2}$. 

%%% BEGIN TABLE %%%
\begin{table*}[t]
\centering
\begin{tabular}{|c|c|} 
\hline
\multicolumn{2}{|c|}{$U(\phi) = \Lambda + \frac{\Gamma}{2} - \frac{M_{2}^{4}}{2M_{*}^{2}}-\frac{3H\bar{M}_{1}^{3}}{2M_{*}^{2}}+\frac{3H\bar{M}_{3}^{3}}{M_{*}^{2}}-\frac{3(\bar{M}_{1}^{3})^{\prime}}{20}+\frac{(\bar{M}_{3}^{3})^{\prime}}{5}-\frac{\bar{M}_{3}^{4}}{M_{*}^{2}}$} \Tstrut \Bstrut \\ 
\hline
\multicolumn{2}{|c|}{$Z(\phi) = \frac{\Gamma}{M_{*}^{4}} - \frac{2M_{2}^{4}}{M_{*}^{6}}-\frac{3H \bar{M}_{1}^{3}}{M_{*}^{6}}+\frac{12H\bar{M}_{3}^{3}}{M_{*}^{6}}+\frac{3(\bar{M}_{1}^{3})^{\prime}}{5M_{*}^{4}}-\frac{4(\bar{M}_{3}^{3})^{\prime}}{5M_{*}^{4}}-\frac{6\bar{M}_{3}^{4}}{M_{*}^{6}} $} \Tstrut \Bstrut  \\ \hline 
\multicolumn{2}{|c|}
{$a_{2}(\phi)=\frac{M_{2}^{4}}{2M_{*}^{8}}-\frac{3H\bar{M}_{3}^{3}}{M_{*}^{8}}+\frac{(\bar{M}_{1}^{3})^{\prime}}{5M_{*}^{6}}-\frac{3(\bar{M}_{3}^{3})^{\prime}}{5M_{*}^{6}}+\frac{3\bar{M}_{3}^{4}}{M_{*}^{8}}$} \Bstrut \Tstrut \\ 
\hline                   
 $ a_{3}(\phi)=\frac{(\bar{M}_{1}^{3})^{\prime}}{40M_{*}^{10}}-\frac{(\bar{M}_{3}^{3})^{\prime}}{5M_{*}^{10}}+\frac{\bar{M}_{3}^{4}}{M_{*}^{12}} $ \hspace{0.1cm} &  $b_{1}(\phi)= \frac{3\bar{M}_{1}^{3}}{4M_{*}^{6}}-\frac{2\bar{M}_{3}^{3}}{M_{*}^{6}}$  \Bstrut \Tstrut  \\ 
\hline 
$F(\phi)=\Omega$ & $b_2(\phi)=\frac{\bar{M}_{1}^{3}}{8M_{*}^{10}}-\frac{\bar{M}_{3}^{3}}{M_{*}^{10}}$ \Bstrut \Tstrut \\ 
\hline
\end{tabular}
\caption{Contributions to the reconstructed Horndeski action arising from the nonlinear corrections in the EFT action at third order. The reconstruction can easily be expanded to arbitrary higher order.
} 
\label{nonlinearsolution}
\end{table*}
%%% END TABLE %%%

Importantly, this method can straightforwardly be extended to higher orders, where at each order it is necessary to invert an $n\times n$ matrix to obtain the corresponding EFT coefficients in terms of covariant functions in $G_{3}$. 
It is then possible to derive a reconstruction from the $\bar{M}_{i}^{4}, \bar{M}_{i}^{3}$ terms which proceeds in exactly the same manner as discussed for $n=3$.
It is also important to stress that a different combination of the terms in Eq.~\eqref{XmBoxphi} with different choices of $m$ could have been chosen to develop the reconstruction. From the structure of Eq.~\eqref{XmBoxphi} there will always be terms involving $\left(\delta g^{00}\right)^{n}\delta K$ to arbitrary order for any $m$ which can be used as the basis for deriving the reconstructed theory.
There is therefore a degeneracy in the space of models which go as $X^{m}\Box \phi$ on the behaviour of the background and perturbations. 

The reconstructed Horndeski theory that covers the background, linear- and second-order cosmological perturbations is given by  
\begin{align}
G_{2}(\phi, X) = & -M_{*}^{2}U(\phi) - \frac{1}{2}M_{*}^{2} Z(\phi)X+a_{2}(\phi)X^{2} \nonumber\\
 &+a_{3}(\phi)X^{3}+\Delta G_{2} \,,
\label{eq:G2reconnonlinear} \\
G_{3}(\phi,X) = & \: b_{0}(\phi)+b_{1}(\phi)X+b_{2}(\phi)X^{2}+\Delta G_{3} \,,
\label{eq:G3reconnonlinear} \\
G_{4}(\phi, X) = & \: \frac{1}{2}M_{*}^{2}F(\phi) \,.
\label{eq:G4reconnonlinear}
\end{align}
The precise form of each term written in terms of the EFT functions is presented in Table~\ref{nonlinearsolution}. Note that now that we have extended the reconstruction to nonlinear order it is necessary to include higher powers of $X$ in the reconstruction, both in $G_{2}$ and $G_{3}$.
In the same manner, if we were to extend the reconstruction to $(n-1)$-th order in cosmological perturbations it would introduce terms of the form $X^{n}$ in $G_{2}$ and $G_{3}$.

Finally, it is also of interest to examine what effect these higher-order perturbations have on the physical EFT basis recently introduced in Refs.~\cite{Kennedy:2018gtx, 2019JCAP...01..041L}. It consists of parameterizing the EFT formalism in terms of inherently stable basis functions: The effective Planck mass squared $M^{2}$, the sound-speed squared $c_{s}^{2}$, the kinetic energy of the scalar field $\alpha$ and the background expansion $H(t)$, along with $\alpha_{B0}$.
Any constraints placed on these parameters are guaranteed to satisfy the conditions for avoiding ghost and gradient instabilities, which otherwise must be checked
independently for other bases.
For higher-order perturbations, note that by
shifting the time coordinate infinitesimally such that $t\rightarrow t+\pi$ the important operators for our purpose in the EFT action change in accordance with the following St{\"u}ckelberg transformations~\cite{Gubitosi2012, Bloomfield:2012ff} 
\begin{equation}
    g^{00}\rightarrow g^{00}+2g^{0\mu}\partial_{\mu}\pi+g^{\mu\nu}\partial_{\mu}\partial_{\nu}\pi \, ,
\end{equation}
\begin{equation}
    \delta K \rightarrow \delta K-3\dot{H}\pi-a^{-2}\Box \pi \, ,
\end{equation}
where $\pi$ is interpreted as the extra scalar degree of freedom which was hidden when the action was written in the unitary gauge. An operator of the form $\bar{M}_{3}^{4}(t)(\delta g^{00})^{3}$ will introduce terms in the full Lagrangian such as $\bar{M}_{3}^{4}(t)\dot{\pi}^{2}$ after applying the time diffeomorphism.
As the physical basis for the EFT functions is defined through the coefficients of such terms, this implies that these higher-order operators act to correct the lower-order EFT functions.
For example, the soundspeed will now depend on these higher-order EFT functions and so the linear stability may be affected by what occurs at the nonlinear level. Physically this makes sense. If one has a second-order perturbation which is unstable, it will produce a runaway effect such that it will grow to affect the linear and background scales. In other words, the perturbations of the perturbations must be kept under control if the theory is to be completely stable. The stability of the full theory can of course be computed at the level of the covariant action.
EFT naturally splits up the dynamics of the different length scales, and in order to obtain a theory that is stable, this stability must be kept at all orders in the EFT expansion.
We leave a discussion of these issues for future work.
%

%%%%%% CONCLUSIONS %%%%%
\section{Conclusions} \label{sec:conclusions}
%%%%%%%%%%%%%%%%%%%%%%%%

Constraining models beyond $\Lambda$CDM is a worthwhile and promising endeavor of
modern cosmology. We are about to see an enormous influx of observational data from surveys such as Euclid~\cite{Laureijs:2011gra, Amendola:2012ys} and LSST~\cite{Ivezic:2008fe}, which will provide percent-level constraints on the cosmological parameters.
The outcome of these surveys will be twofold.
Either the Universe turns out to be consistent with $\Lambda$CDM, which will motivate a more directed effort in tackling the cosmological constant problem (see, e.g.,~Refs.~\cite{Barrow:2010xt,Shaw:2010pq,Barrow:2011zp,Casadio:2013uia,Kaloper:2013zca,Kaloper:2015jra,Wang:2017oiy,Lombriser:2017cjy,Appleby:2018yci,Lombriser:2018aru,Wang:2018kly,Evnin:2018zeo,Canales:2018tbn,Lombriser:2019jia}).
On the other hand, if recent observational tensions~\cite{Hildebrandt:2016iqg, 2018ApJ...855..136R, Aghanim:2018eyx} persist then that will be strong evidence that the theory describing the Universe on cosmological scales requires revision and potentially will go beyond a cosmological constant.
Constraints on deviations from GR are obtained on a broad range of different length scales, and a potentially new theory on large cosmological scalesmust also be consistent with observations at nonlinear scales.
In this paper we have discussed how in generalised scalar-tensor theories observations made at the level of the background and the linear perturbations may be connected with the nonlinear regime and vice-versa.
This is made possible through the reconstruction of covariant Horndeski theory from the EFT of dark energy~\cite{Kennedy:2017sof, Kennedy:2018gtx}.
The reconstructed theories are degenerate to linear order in cosmological perturbations and differ only by nonlinear correction terms $\Delta G_{i}$. 
We first explored the uniqueness of these corrections terms.
At $n$-th order in perturbation theory the number of EFT operators that one can write down which are consistent with the symmetry of broken time diffeomorphisms becomes unmanageable.
However, we have argued that by restricting to Horndeski theories that respect the GW170817 constraint of luminal speed of gravity~\cite{Monitor:2017mdv, McManus:2016kxu} the number of free functions that enter the EFT expansion at each order is limited to two.
The two correction terms at $n$-th order can then be related to the free functions $\xi^{{\scriptscriptstyle(2,3)}}_{n}(\phi)$ specifying $\Delta G_{2}$ and $\Delta G_{3}$.

As a first application of the nonlinear correction terms, we have considered the implementation of screening mechanisms.
%%%%%%%%%%%%%%%%Screening%%%%%%%%%%%%%%%%%%%%
With the reconstructed covariant theory it is possible to apply techniques that have been developed~\cite{McManus:2016kxu, McManus:2017itv} to
identify the existence of Einstein gravity limits
within a given Horndeski theory.
With the use of these methods we have demonstrated that there is enough freedom on nonlinear scales to
employ a particular type of a screening mechanism by a suitable configuration of the correction terms.
More specifically, we have provided the examples of realizing a chameleon, k-mouflage and Vanshtein mechanism.
%

%%%%%%%%%%%Kinetic SA%%%%%%%%%%%%%%%%%%
A further consequence of the reconstruction method concerns the identification of a class of models that is degenerate with $\Lambda$CDM at the level of the background and linear perturbations but departs from it at arbitrary order of nonlinear perturbations.
A subclass of these models further exhibits kinetic self-acceleration, where the background expansion is accelerating exactly like $\Lambda$CDM but there is no explicit cosmological constant written in the theory. The acceleration is instead driven by the kinetic terms.
An immediate consequence of the existence of such models is that even if the background expansion and linear matter power spectrum is measured to agree with $\Lambda$CDM from the next generation of surveys, the degenerate alternatives may not generally be excluded.
Moreover, a theoretically appealing aspect of these models is that, with the cosmological constant now acting as a coefficient of kinetic terms rather than a bare constant, it may be possible to render it technically natural.
These implications warrant a more detailed study of these models. 
Finally, the same techniques that were employed in the development of the reconstruction of the Horndeski action to linear order in cosmological perturbations were utilized here to derive a reconstructed theory that includes the nonlinear EFT functions.
For given constraints on these functions this enables a reconstruction of the Horndeski theory across a broad range of length scales, which may be supplemented with a restriction of the allowed forms of $\Delta G_{i}$ to those that employ a screening mechanism.

There remain many further applications to be examined for the nonlinear sector of the reconstruction method. For example, obtaining the stability conditions is an important step in understanding the viability of the sampled models in parameter estimation analyses and it is as yet unclear what effect the nonlinear correction terms have on the stability of the theory.
There may also be a more physical basis for the correction terms such as that presented in Ref.~\cite{2019JCAP...01..041L} for linear perturbations, which automatically satisfies the stability constraints at the nonlinear level. We leave such considerations to upcoming studies.

%%%% ACKNOWLEDGMENTS %%%
\acknowledgments
%%%%%%%%%%%%%%%%%%%%%%%%
%
We thank Ryan McManus for useful discussions.
This work was supported by the STFC Consolidated Grant for Astronomy and Astrophysics at the University of Edinburgh.
J.K. thanks STFC for support through an STFC studentship.
L.L. acknowledges support by a Swiss National Science Foundation Professorship grant (No.~170547) and the Affiliate programme of the Higgs Centre for Theoretical Physics. A.N.T. thanks the Royal Society for support from a Wolfson Research Merit Award.
Please contact the authors for access to research materials.

\appendix 

\section{Horndeski field equations with $\alpha_{T}=0$}
\label{appendix}

For completeness, we shall present here the metric and scalar field equations that are obtained from varying $g_{\mu\nu}$ and $\phi$ in Eqs.~\eqref{eq:HorndeskiL2at0} to \eqref{eq:g4at0}. Although the structure of these equations is complicated the relevance for the application in Sec.~\ref{sec:screening} is simply the number of spacetime derivatives and powers of the scalar field that enter into each of the field equations. The metric field equation is given by~\cite{McManus:2016kxu}
\begin{equation}
    \Gamma R_{\mu\nu}=-\sum_{i=2}^{4}\mathcal{T}^{(i)}_{\mu\nu}+\left(T_{\mu\nu}-\frac{1}{2}g_{\mu\nu}T \right)/M_{*}^{2}
    \label{MFequation}
\end{equation}
and the scalar field equation is given by
\begin{equation}
\Gamma\sum_{i=2,3,4}(\nabla^{\mu}J_{\mu}^{(i)}-P_{\phi}^{(i)}) 
+\Xi \sum_{i=2}^{4} \mathcal{T}^{(i)}=-\frac{T}{M_{*}^{2}} \Xi \, ,
\label{SFequation}
\end{equation}
where $\Gamma \equiv 2G_{4}/M_{*}^{2}$ and $\Xi \equiv 2G_{4\phi}/M_{*}^{2}$ and
\begin{align}
 P_{\phi}^{(2)}=&\frac{2}{M_{*}^{2}}G_{2\phi} \, , \label{Eq1} \\
 P_{\phi}^{(3)}=&\frac{2}{M_{*}^{2}}\nabla_{\mu}G_{3\phi}\nabla^{\mu}\phi \,, \label{Eq2} \\
  P_{\phi}^{(4)}=&\frac{2}{M_{*}^{2}}G_{4\phi}R \,, \label{Eq3} \\
  J^{(2)}_{\mu}=&-G_{2X}\nabla_{\mu}\phi \, ,
  \label{Eq4} \\
    J^{(3)}_{\mu}=&-G_{3X}\Box \phi\nabla_{\mu}\phi+G_{3X}\nabla_{\mu}X+2G_{3\phi}\nabla_{\mu}\phi \, , \label{Eq5} \\
     \mathcal{T}^{(2)}_{\mu\nu}=&-\frac{1}{M_{*}^{2}}G_{2X}\nabla_{\mu}\phi\nabla_{\nu}\phi \nonumber \\ 
&+\frac{1}{2M_{*}^{2}}g_{\mu\nu}\left(XG_{2X}+2G_{2} \right) \, , \label{Eq6} \\
 \mathcal{T}^{(3)}_{\mu\nu}=&\frac{2}{M_{*}^{2}}G_{3X}\mathcal{S}^{(4,3)}+G_{3\phi}\nabla_{\mu}\phi\nabla_{\nu}\phi \, , \label{Eq7} \\
 \mathcal{T}^{(4)}_{\mu\nu}=&G_{4\phi}\mathcal{S}^{(2,1)}+G_{4\phi\phi}\mathcal{S}^{(2,2)} \, . \label{Eq8}
\end{align}
Note that $J_{\mu}^{(4)}=0$. The $\mathcal{S}^{(i,j)}$ notation indicates a term that contains $i$ spacetime derivatives and $j$ powers of the scalar field.
As discussed in Sec.~\ref{sec:screening}, knowledge of these quantities is sufficient to determine whether a given term
will become dominant or sub-dominant in a screened or un-screened limit, not its precise functional form. We refer the reader to the appendix of Ref.~\cite{McManus:2016kxu} for the explicit expressions but note the different definitions of the $G_{i}$ functions and $X$.

%%%%%% BIBLIOGRAHY %%%%%
\bibliographystyle{JHEP}
\bibliography{library}

\providecommand{\href}[2]{#2}\begingroup\raggedright\begin{thebibliography}{10}

\bibitem{Perlmutter:1998np}
{\scshape Supernova Cosmology Project} collaboration, S.~Perlmutter et~al.,
  \emph{{Measurements of Omega and Lambda from 42 high redshift supernovae}},
  \href{http://dx.doi.org/10.1086/307221}{\emph{Astrophys. J.} {\bfseries 517}
  (1999) 565--586}, [\href{https://arxiv.org/abs/astro-ph/9812133}{{\ttfamily
  astro-ph/9812133}}].

\bibitem{Riess:1998cb}
{\scshape Supernova Search Team} collaboration, A.~G. Riess et~al.,
  \emph{{Observational evidence from supernovae for an accelerating universe
  and a cosmological constant}},
  \href{http://dx.doi.org/10.1086/300499}{\emph{Astron. J.} {\bfseries 116}
  (1998) 1009--1038}, [\href{https://arxiv.org/abs/astro-ph/9805201}{{\ttfamily
  astro-ph/9805201}}].

\bibitem{Aghanim:2018eyx}
{\scshape Planck} collaboration, N.~Aghanim et~al., \emph{{Planck 2018 results.
  VI. Cosmological parameters}},
  \href{https://arxiv.org/abs/1807.06209}{{\ttfamily 1807.06209}}.

\bibitem{Martin:2012bt}
J.~Martin, \emph{{Everything You Always Wanted To Know About The Cosmological
  Constant Problem (But Were Afraid To Ask)}},
  \href{http://dx.doi.org/10.1016/j.crhy.2012.04.008}{\emph{Comptes Rendus
  Physique} {\bfseries 13} (2012) 566--665},
  [\href{https://arxiv.org/abs/1205.3365}{{\ttfamily 1205.3365}}].

\bibitem{Weinberg1989}
S.~Weinberg, \emph{{The cosmological constant problem}}, {\emph{Rev. Mod.
  Phys.} {\bfseries 61} (1989) }.

\bibitem{Clifton:2011jh}
T.~Clifton, P.~G. Ferreira, A.~Padilla and C.~Skordis, \emph{{Modified Gravity
  and Cosmology}},
  \href{http://dx.doi.org/10.1016/j.physrep.2012.01.001}{\emph{Phys. Rept.}
  {\bfseries 513} (2012) 1--189},
  [\href{https://arxiv.org/abs/1106.2476}{{\ttfamily 1106.2476}}].

\bibitem{Joyce:2014kja}
A.~Joyce, B.~Jain, J.~Khoury and M.~Trodden, \emph{{Beyond the Cosmological
  Standard Model}},
  \href{http://dx.doi.org/10.1016/j.physrep.2014.12.002}{\emph{Phys. Rept.}
  {\bfseries 568} (2015) 1--98},
  [\href{https://arxiv.org/abs/1407.0059}{{\ttfamily 1407.0059}}].

\bibitem{Joyce:2016vqv}
A.~Joyce, L.~Lombriser and F.~Schmidt, \emph{{Dark Energy Versus Modified
  Gravity}},
  \href{http://dx.doi.org/10.1146/annurev-nucl-102115-044553}{\emph{Ann. Rev.
  Nucl. Part. Sci.} {\bfseries 66} (2016) 95--122},
  [\href{https://arxiv.org/abs/1601.06133}{{\ttfamily 1601.06133}}].

\bibitem{Woodard:2006nt}
R.~P. Woodard, \emph{{Avoiding dark energy with 1/r modifications of gravity}},
  \href{http://dx.doi.org/10.1007/978-3-540-71013-4_14}{\emph{Lect. Notes
  Phys.} {\bfseries 720} (2007) 403--433},
  [\href{https://arxiv.org/abs/astro-ph/0601672}{{\ttfamily
  astro-ph/0601672}}].

\bibitem{Horndeski:1974wa}
G.~W. Horndeski, \emph{{Second-order scalar-tensor field equations in a
  four-dimensional space}},
  \href{http://dx.doi.org/10.1007/BF01807638}{\emph{Int. J. Theor. Phys.}
  {\bfseries 10} (1974) 363--384}.

\bibitem{Deffayet:2011gz}
C.~Deffayet, X.~Gao, D.~A. Steer and G.~Zahariade, \emph{{From k-essence to
  generalised Galileons}},
  \href{http://dx.doi.org/10.1103/PhysRevD.84.064039}{\emph{Phys. Rev.}
  {\bfseries D84} (2011) 064039},
  [\href{https://arxiv.org/abs/1103.3260}{{\ttfamily 1103.3260}}].

\bibitem{Kobayashi2011}
T.~Kobayashi, M.~Yamaguchi and J.~Yokoyama, \emph{{Generalized G-Inflation
  �Inflation with the Most General Second-Order Field Equations�}},
  \href{http://dx.doi.org/10.1143/PTP.126.511}{\emph{Progress of Theoretical
  Physics} {\bfseries 126} (2011) 511--529},
  [\href{https://arxiv.org/abs/1105.5723}{{\ttfamily 1105.5723}}].

\bibitem{Gleyzes:2014dya}
J.~Gleyzes, D.~Langlois, F.~Piazza and F.~Vernizzi, \emph{{Healthy theories
  beyond Horndeski}},
  \href{http://dx.doi.org/10.1103/PhysRevLett.114.211101}{\emph{Phys. Rev.
  Lett.} {\bfseries 114} (2015) 211101},
  [\href{https://arxiv.org/abs/1404.6495}{{\ttfamily 1404.6495}}].

\bibitem{Langlois:2015cwa}
D.~Langlois and K.~Noui, \emph{{Degenerate higher derivative theories beyond
  Horndeski: evading the Ostrogradski instability}},
  \href{http://dx.doi.org/10.1088/1475-7516/2016/02/034}{\emph{JCAP} {\bfseries
  1602} (2016) 034}, [\href{https://arxiv.org/abs/1510.06930}{{\ttfamily
  1510.06930}}].

\bibitem{Weinberg:2008hq}
S.~Weinberg, \emph{{Effective Field Theory for Inflation}},
  \href{http://dx.doi.org/10.1103/PhysRevD.77.123541}{\emph{Phys. Rev.}
  {\bfseries D77} (2008) 123541},
  [\href{https://arxiv.org/abs/0804.4291}{{\ttfamily 0804.4291}}].

\bibitem{Cheung:2007st}
C.~Cheung, P.~Creminelli, A.~L. Fitzpatrick, J.~Kaplan and L.~Senatore,
  \emph{{The Effective Field Theory of Inflation}},
  \href{http://dx.doi.org/10.1088/1126-6708/2008/03/014}{\emph{JHEP} {\bfseries
  03} (2008) 014}, [\href{https://arxiv.org/abs/0709.0293}{{\ttfamily
  0709.0293}}].

\bibitem{Creminelli:2008wc}
P.~Creminelli, G.~D'Amico, J.~Norena and F.~Vernizzi, \emph{{The Effective
  Theory of Quintessence: the w\ $<$\,-1 Side Unveiled}},
  \href{http://dx.doi.org/10.1088/1475-7516/2009/02/018}{\emph{JCAP} {\bfseries
  0902} (2009) 018}, [\href{https://arxiv.org/abs/0811.0827}{{\ttfamily
  0811.0827}}].

\bibitem{Park:2010cw}
M.~Park, K.~M. Zurek and S.~Watson, \emph{{A Unified Approach to Cosmic
  Acceleration}},
  \href{http://dx.doi.org/10.1103/PhysRevD.81.124008}{\emph{Phys. Rev.}
  {\bfseries D81} (2010) 124008},
  [\href{https://arxiv.org/abs/1003.1722}{{\ttfamily 1003.1722}}].

\bibitem{Bloomfield:2011np}
J.~K. Bloomfield and E.~E. Flanagan, \emph{{A Class of Effective Field Theory
  Models of Cosmic Acceleration}},
  \href{http://dx.doi.org/10.1088/1475-7516/2012/10/039}{\emph{JCAP} {\bfseries
  1210} (2012) 039}, [\href{https://arxiv.org/abs/1112.0303}{{\ttfamily
  1112.0303}}].

\bibitem{Gubitosi2012}
G.~Gubitosi, F.~Piazza and F.~Vernizzi, \emph{{The Effective Field Theory of
  Dark Energy}},
  \href{http://dx.doi.org/10.1088/1475-7516/2013/02/032}{\emph{JCAP} {\bfseries
  1302} (2013) 032}, [\href{https://arxiv.org/abs/1210.0201}{{\ttfamily
  1210.0201}}].

\bibitem{Bloomfield:2012ff}
J.~K. Bloomfield, E.~E. Flanagan, M.~Park and S.~Watson, \emph{{Dark energy or
  modified gravity? An effective field theory approach}},
  \href{http://dx.doi.org/10.1088/1475-7516/2013/08/010}{\emph{JCAP} {\bfseries
  1308} (2013) 010}, [\href{https://arxiv.org/abs/1211.7054}{{\ttfamily
  1211.7054}}].

\bibitem{Gleyzes:2013ooa}
J.~Gleyzes, D.~Langlois, F.~Piazza and F.~Vernizzi, \emph{{Essential Building
  Blocks of Dark Energy}},
  \href{http://dx.doi.org/10.1088/1475-7516/2013/08/025}{\emph{JCAP} {\bfseries
  1308} (2013) 025}, [\href{https://arxiv.org/abs/1304.4840}{{\ttfamily
  1304.4840}}].

\bibitem{Bloomfield:2013efa}
J.~Bloomfield, \emph{{A Simplified Approach to General Scalar-Tensor
  Theories}},
  \href{http://dx.doi.org/10.1088/1475-7516/2013/12/044}{\emph{JCAP} {\bfseries
  1312} (2013) 044}, [\href{https://arxiv.org/abs/1304.6712}{{\ttfamily
  1304.6712}}].

\bibitem{Tsujikawa:2014mba}
S.~Tsujikawa, \emph{{The effective field theory of inflation/dark energy and
  the Horndeski theory}},
  \href{http://dx.doi.org/10.1007/978-3-319-10070-8_4}{\emph{Lect. Notes Phys.}
  {\bfseries 892} (2015) 97--136},
  [\href{https://arxiv.org/abs/1404.2684}{{\ttfamily 1404.2684}}].

\bibitem{Gleyzes:2014rba}
J.~Gleyzes, D.~Langlois and F.~Vernizzi, \emph{{A unifying description of dark
  energy}}, \href{http://dx.doi.org/10.1142/S021827181443010X}{\emph{Int. J.
  Mod. Phys.} {\bfseries D23} (2015) 1443010},
  [\href{https://arxiv.org/abs/1411.3712}{{\ttfamily 1411.3712}}].

\bibitem{Bellini:2014fua}
E.~Bellini and I.~Sawicki, \emph{{Maximal freedom at minimum cost: linear
  large-scale structure in general modifications of gravity}},
  \href{http://dx.doi.org/10.1088/1475-7516/2014/07/050}{\emph{JCAP} {\bfseries
  1407} (2014) 050}, [\href{https://arxiv.org/abs/1404.3713}{{\ttfamily
  1404.3713}}].

\bibitem{Lagos:2016wyv}
M.~Lagos, T.~Baker, P.~G. Ferreira and J.~Noller, \emph{{A general theory of
  linear cosmological perturbations: scalar-tensor and vector-tensor
  theories}},
  \href{http://dx.doi.org/10.1088/1475-7516/2016/08/007}{\emph{JCAP} {\bfseries
  1608} (2016) 007}, [\href{https://arxiv.org/abs/1604.01396}{{\ttfamily
  1604.01396}}].

\bibitem{Cusin:2017mzw}
G.~Cusin, M.~Lewandowski and F.~Vernizzi, \emph{{Nonlinear Effective Theory of
  Dark Energy}},
  \href{http://dx.doi.org/10.1088/1475-7516/2018/04/061}{\emph{JCAP} {\bfseries
  1804} (2018) 061}, [\href{https://arxiv.org/abs/1712.02782}{{\ttfamily
  1712.02782}}].

\bibitem{Frusciante:2017nfr}
N.~Frusciante and G.~Papadomanolakis, \emph{{Tackling non-linearities with the
  effective field theory of dark energy and modified gravity}},
  \href{http://dx.doi.org/10.1088/1475-7516/2017/12/014}{\emph{JCAP} {\bfseries
  1712} (2017) 014}, [\href{https://arxiv.org/abs/1706.02719}{{\ttfamily
  1706.02719}}].

\bibitem{Kennedy:2017sof}
J.~Kennedy, L.~Lombriser and A.~Taylor, \emph{{Reconstructing Horndeski models
  from the effective field theory of dark energy}},
  \href{http://dx.doi.org/10.1103/PhysRevD.96.084051}{\emph{Phys. Rev.}
  {\bfseries D96} (2017) 084051},
  [\href{https://arxiv.org/abs/1705.09290}{{\ttfamily 1705.09290}}].

\bibitem{Kennedy:2018gtx}
J.~Kennedy, L.~Lombriser and A.~Taylor, \emph{{Reconstructing Horndeski
  theories from phenomenological modified gravity and dark energy models on
  cosmological scales}},
  \href{http://dx.doi.org/10.1103/PhysRevD.98.044051}{\emph{Phys. Rev.}
  {\bfseries D98} (2018) 044051},
  [\href{https://arxiv.org/abs/1804.04582}{{\ttfamily 1804.04582}}].

\bibitem{Monitor:2017mdv}
{\scshape Virgo, Fermi-GBM, INTEGRAL, LIGO Scientific} collaboration, B.~P.
  Abbott et~al., \emph{{Gravitational Waves and Gamma-rays from a Binary
  Neutron Star Merger: GW170817 and GRB 170817A}},
  \href{http://dx.doi.org/10.3847/2041-8213/aa920c}{\emph{Astrophys. J.}
  {\bfseries 848} (2017) L13},
  [\href{https://arxiv.org/abs/1710.05834}{{\ttfamily 1710.05834}}].

\bibitem{Bellini:2015wfa}
E.~Bellini, R.~Jimenez and L.~Verde, \emph{{Signatures of Horndeski gravity on
  the Dark Matter Bispectrum}},
  \href{http://dx.doi.org/10.1088/1475-7516/2015/05/057}{\emph{JCAP} {\bfseries
  1505} (2015) 057}, [\href{https://arxiv.org/abs/1504.04341}{{\ttfamily
  1504.04341}}].

\bibitem{Will:2014kxa}
C.~M. Will, \emph{{The Confrontation between General Relativity and
  Experiment}}, \href{http://dx.doi.org/10.12942/lrr-2014-4}{\emph{Living Rev.
  Rel.} {\bfseries 17} (2014) 4},
  [\href{https://arxiv.org/abs/1403.7377}{{\ttfamily 1403.7377}}].

\bibitem{Khoury:2003aq}
J.~Khoury and A.~Weltman, \emph{{Chameleon fields: Awaiting surprises for tests
  of gravity in space}},
  \href{http://dx.doi.org/10.1103/PhysRevLett.93.171104}{\emph{Phys. Rev.
  Lett.} {\bfseries 93} (2004) 171104},
  [\href{https://arxiv.org/abs/astro-ph/0309300}{{\ttfamily
  astro-ph/0309300}}].

\bibitem{Hinterbichler:2010es}
K.~Hinterbichler and J.~Khoury, \emph{{Symmetron Fields: Screening Long-Range
  Forces Through Local Symmetry Restoration}},
  \href{http://dx.doi.org/10.1103/PhysRevLett.104.231301}{\emph{Phys. Rev.
  Lett.} {\bfseries 104} (2010) 231301},
  [\href{https://arxiv.org/abs/1001.4525}{{\ttfamily 1001.4525}}].

\bibitem{Babichev:2009ee}
E.~Babichev, C.~Deffayet and R.~Ziour, \emph{{k-Mouflage gravity}},
  \href{http://dx.doi.org/10.1142/S0218271809016107}{\emph{Int. J. Mod. Phys.}
  {\bfseries D18} (2009) 2147--2154},
  [\href{https://arxiv.org/abs/0905.2943}{{\ttfamily 0905.2943}}].

\bibitem{Vainshtein:1972sx}
A.~I. Vainshtein, \emph{{To the problem of nonvanishing gravitation mass}},
  \href{http://dx.doi.org/10.1016/0370-2693(72)90147-5}{\emph{Phys. Lett.}
  {\bfseries 39B} (1972) 393--394}.

\bibitem{McManus:2016kxu}
R.~McManus, L.~Lombriser and J.~Peñarrubia, \emph{{Finding Horndeski theories
  with Einstein gravity limits}},
  \href{http://dx.doi.org/10.1088/1475-7516/2016/11/006}{\emph{JCAP} {\bfseries
  1611} (2016) 006}, [\href{https://arxiv.org/abs/1606.03282}{{\ttfamily
  1606.03282}}].

\bibitem{McManus:2017itv}
R.~McManus, L.~Lombriser and J.~Peñarrubia, \emph{{Parameterised
  Post-Newtonian Expansion in Screened Regions}},
  \href{http://dx.doi.org/10.1088/1475-7516/2017/12/031}{\emph{JCAP} {\bfseries
  1712} (2017) 031}, [\href{https://arxiv.org/abs/1705.05324}{{\ttfamily
  1705.05324}}].

\bibitem{Lombriser:2014ira}
L.~Lombriser and A.~Taylor, \emph{{Classifying Linearly Shielded Modified
  Gravity Models in Effective Field Theory}},
  \href{http://dx.doi.org/10.1103/PhysRevLett.114.031101}{\emph{Phys. Rev.
  Lett.} {\bfseries 114} (2015) 031101},
  [\href{https://arxiv.org/abs/1405.2896}{{\ttfamily 1405.2896}}].

\bibitem{Lombriser:2015sxa}
L.~Lombriser and A.~Taylor, \emph{{Breaking a Dark Degeneracy with
  Gravitational Waves}},
  \href{http://dx.doi.org/10.1088/1475-7516/2016/03/031}{\emph{JCAP} {\bfseries
  1603} (2016) 031}, [\href{https://arxiv.org/abs/1509.08458}{{\ttfamily
  1509.08458}}].

\bibitem{Lombriser:2016yzn}
L.~Lombriser and N.~A. Lima, \emph{{Challenges to Self-Acceleration in Modified
  Gravity from Gravitational Waves and Large-Scale Structure}},
  \href{http://dx.doi.org/10.1016/j.physletb.2016.12.048}{\emph{Phys. Lett.}
  {\bfseries B765} (2017) 382--385},
  [\href{https://arxiv.org/abs/1602.07670}{{\ttfamily 1602.07670}}].

\bibitem{2019JCAP...01..041L}
L.~{Lombriser}, C.~{Dalang}, J.~{Kennedy} and A.~{Taylor}, \emph{{Inherently
  stable effective field theory for dark energy and modified gravity}},
  \href{http://dx.doi.org/10.1088/1475-7516/2019/01/041}{\emph{JCAP} {\bfseries
  0119} (2019) 041}, [\href{https://arxiv.org/abs/1810.05225}{{\ttfamily
  1810.05225}}].

\bibitem{Chow:2009fm}
N.~Chow and J.~Khoury, \emph{{Galileon Cosmology}},
  \href{http://dx.doi.org/10.1103/PhysRevD.80.024037}{\emph{Phys. Rev.}
  {\bfseries D80} (2009) 024037},
  [\href{https://arxiv.org/abs/0905.1325}{{\ttfamily 0905.1325}}].

\bibitem{PhysRevD.80.121301}
F.~P. Silva and K.~Koyama, \emph{Self-accelerating universe in galileon
  cosmology}, \href{http://dx.doi.org/10.1103/PhysRevD.80.121301}{\emph{Phys.
  Rev. D} {\bfseries 80} (Dec, 2009) 121301}.

\bibitem{Caldwell:1997ii}
R.~R. Caldwell, R.~Dave and P.~J. Steinhardt, \emph{{Cosmological imprint of an
  energy component with general equation of state}},
  \href{http://dx.doi.org/10.1103/PhysRevLett.80.1582}{\emph{Phys. Rev. Lett.}
  {\bfseries 80} (1998) 1582--1585},
  [\href{https://arxiv.org/abs/astro-ph/9708069}{{\ttfamily
  astro-ph/9708069}}].

\bibitem{PhysRevD.62.023511}
T.~Chiba, T.~Okabe and M.~Yamaguchi, \emph{Kinetically driven quintessence},
  \href{http://dx.doi.org/10.1103/PhysRevD.62.023511}{\emph{Phys. Rev. D}
  {\bfseries 62} (Jun, 2000) 023511}.

\bibitem{ArmendarizPicon:2000ah}
C.~Armendariz-Picon, V.~F. Mukhanov and P.~J. Steinhardt, \emph{{Essentials of
  k essence}}, \href{http://dx.doi.org/10.1103/PhysRevD.63.103510}{\emph{Phys.
  Rev.} {\bfseries D63} (2001) 103510},
  [\href{https://arxiv.org/abs/astro-ph/0006373}{{\ttfamily
  astro-ph/0006373}}].

\bibitem{Lombriser2015}
L.~Lombriser and A.~Taylor, \emph{{Semi-dynamical perturbations of unified dark
  energy}},  \href{https://arxiv.org/abs/1505.05915}{{\ttfamily 1505.05915}}.

\bibitem{Uzan:2006mf}
J.-P. Uzan, \emph{{The acceleration of the universe and the physics behind
  it}}, \href{http://dx.doi.org/10.1007/s10714-006-0385-z}{\emph{Gen. Rel.
  Grav.} {\bfseries 39} (2007) 307--342},
  [\href{https://arxiv.org/abs/astro-ph/0605313}{{\ttfamily
  astro-ph/0605313}}].

\bibitem{Amendola:2007rr}
L.~Amendola, M.~Kunz and D.~Sapone, \emph{{Measuring the dark side (with weak
  lensing)}},
  \href{http://dx.doi.org/10.1088/1475-7516/2008/04/013}{\emph{JCAP} {\bfseries
  0804} (2008) 013}, [\href{https://arxiv.org/abs/0704.2421}{{\ttfamily
  0704.2421}}].

\bibitem{Caldwell:2007cw}
R.~Caldwell, A.~Cooray and A.~Melchiorri, \emph{{Constraints on a New
  Post-General Relativity Cosmological Parameter}},
  \href{http://dx.doi.org/10.1103/PhysRevD.76.023507}{\emph{Phys. Rev.}
  {\bfseries D76} (2007) 023507},
  [\href{https://arxiv.org/abs/astro-ph/0703375}{{\ttfamily
  astro-ph/0703375}}].

\bibitem{Hu:2007pj}
W.~Hu and I.~Sawicki, \emph{{A Parameterized Post-Friedmann Framework for
  Modified Gravity}},
  \href{http://dx.doi.org/10.1103/PhysRevD.76.104043}{\emph{Phys. Rev.}
  {\bfseries D76} (2007) 104043},
  [\href{https://arxiv.org/abs/0708.1190}{{\ttfamily 0708.1190}}].

\bibitem{Zhang:2007nk}
P.~Zhang, M.~Liguori, R.~Bean and S.~Dodelson, \emph{{Probing Gravity at
  Cosmological Scales by Measurements which Test the Relationship between
  Gravitational Lensing and Matter Overdensity}},
  \href{http://dx.doi.org/10.1103/PhysRevLett.99.141302}{\emph{Phys. Rev.
  Lett.} {\bfseries 99} (2007) 141302},
  [\href{https://arxiv.org/abs/0704.1932}{{\ttfamily 0704.1932}}].

\bibitem{Ade:2015rim}
{\scshape Planck} collaboration, P.~A.~R. Ade et~al., \emph{{Planck 2015
  results. XIV. Dark energy and modified gravity}},
  \href{http://dx.doi.org/10.1051/0004-6361/201525814}{\emph{Astron.
  Astrophys.} {\bfseries 594} (2016) A14},
  [\href{https://arxiv.org/abs/1502.01590}{{\ttfamily 1502.01590}}].

\bibitem{Lombriser:2016zfz}
L.~Lombriser, \emph{{A parametrisation of modified gravity on nonlinear
  cosmological scales}},
  \href{http://dx.doi.org/10.1088/1475-7516/2016/11/039}{\emph{JCAP} {\bfseries
  1611} (2016) 039}, [\href{https://arxiv.org/abs/1608.00522}{{\ttfamily
  1608.00522}}].

\bibitem{2018IJMPD..2748002L}
L.~{Lombriser}, \emph{{Parametrizations for tests of gravity}},
  \href{http://dx.doi.org/10.1142/S0218271818480024}{\emph{International
  Journal of Modern Physics D} {\bfseries 27} (2018) 1848002}.

\bibitem{Bolis:2018kcq}
N.~Bolis, C.~Skordis, D.~B. Thomas and T.~Zlosnik, \emph{{The Parameterized
  Post-Newtonian-Vainshteinian formalism for the Galileon field}},
  \href{https://arxiv.org/abs/1810.02725}{{\ttfamily 1810.02725}}.

\bibitem{PhysRevD.69.044026}
J.~Khoury and A.~Weltman, \emph{Chameleon cosmology},
  \href{http://dx.doi.org/10.1103/PhysRevD.69.044026}{\emph{Phys. Rev. D}
  {\bfseries 69} (Feb, 2004) 044026}.

\bibitem{Lombriser:2013eza}
L.~Lombriser, K.~Koyama and B.~Li, \emph{{Halo modelling in chameleon
  theories}},
  \href{http://dx.doi.org/10.1088/1475-7516/2014/03/021}{\emph{JCAP} {\bfseries
  1403} (2014) 021}, [\href{https://arxiv.org/abs/1312.1292}{{\ttfamily
  1312.1292}}].

\bibitem{Brax:2012jr}
P.~Brax, C.~Burrage and A.-C. Davis, \emph{{Screening fifth forces in k-essence
  and DBI models}},
  \href{http://dx.doi.org/10.1088/1475-7516/2013/01/020}{\emph{JCAP} {\bfseries
  1301} (2013) 020}, [\href{https://arxiv.org/abs/1209.1293}{{\ttfamily
  1209.1293}}].

\bibitem{Cordero:2016bxt}
R.~Cordero, E.~L. Gonzalez and A.~Queijeiro, \emph{{An equation of state for
  purely kinetic k-essence inspired by cosmic topological defects}},
  \href{http://dx.doi.org/10.1140/epjc/s10052-017-4913-7}{\emph{Eur. Phys. J.}
  {\bfseries C77} (2017) 413},
  [\href{https://arxiv.org/abs/1608.06540}{{\ttfamily 1608.06540}}].

\bibitem{Kimura:2011td}
R.~Kimura, T.~Kobayashi and K.~Yamamoto, \emph{{Observational Constraints on
  Kinetic Gravity Braiding from the Integrated Sachs-Wolfe Effect}},
  \href{http://dx.doi.org/10.1103/PhysRevD.85.123503}{\emph{Phys. Rev.}
  {\bfseries D85} (2012) 123503},
  [\href{https://arxiv.org/abs/1110.3598}{{\ttfamily 1110.3598}}].

\bibitem{Gripaios:2015qya}
B.~Gripaios, \emph{{Lectures on Effective Field Theory}},
  \href{https://arxiv.org/abs/1506.05039}{{\ttfamily 1506.05039}}.

\bibitem{Pirtskhalava:2015nla}
D.~Pirtskhalava, L.~Santoni, E.~Trincherini and F.~Vernizzi, \emph{{Weakly
  Broken Galileon Symmetry}},
  \href{http://dx.doi.org/10.1088/1475-7516/2015/09/007}{\emph{JCAP} {\bfseries
  1509} (2015) 007}, [\href{https://arxiv.org/abs/1505.00007}{{\ttfamily
  1505.00007}}].

\bibitem{Laureijs:2011gra}
{\scshape EUCLID} collaboration, R.~Laureijs et~al., \emph{{Euclid Definition
  Study Report}},  \href{https://arxiv.org/abs/1110.3193}{{\ttfamily
  1110.3193}}.

\bibitem{Amendola:2012ys}
{\scshape Euclid Theory Working Group} collaboration, L.~Amendola et~al.,
  \emph{{Cosmology and fundamental physics with the Euclid satellite}},
  \href{http://dx.doi.org/10.12942/lrr-2013-6}{\emph{Living Rev. Rel.}
  {\bfseries 16} (2013) 6}, [\href{https://arxiv.org/abs/1206.1225}{{\ttfamily
  1206.1225}}].

\bibitem{Ivezic:2008fe}
{\scshape LSST} collaboration, Z.~Ivezic, J.~A. Tyson, R.~Allsman, J.~Andrew
  and R.~Angel, \emph{{LSST: from Science Drivers to Reference Design and
  Anticipated Data Products}},
  \href{https://arxiv.org/abs/0805.2366}{{\ttfamily 0805.2366}}.

\bibitem{Barrow:2010xt}
J.~D. Barrow and D.~J. Shaw, \emph{{A New Solution of The Cosmological Constant
  Problems}},
  \href{http://dx.doi.org/10.1103/PhysRevLett.106.101302}{\emph{Phys. Rev.
  Lett.} {\bfseries 106} (2011) 101302},
  [\href{https://arxiv.org/abs/1007.3086}{{\ttfamily 1007.3086}}].

\bibitem{Shaw:2010pq}
D.~J. Shaw and J.~D. Barrow, \emph{{A Testable Solution of the Cosmological
  Constant and Coincidence Problems}},
  \href{http://dx.doi.org/10.1103/PhysRevD.83.043518}{\emph{Phys. Rev.}
  {\bfseries D83} (2011) 043518},
  [\href{https://arxiv.org/abs/1010.4262}{{\ttfamily 1010.4262}}].

\bibitem{Barrow:2011zp}
J.~D. Barrow and D.~J. Shaw, \emph{{The Value of the Cosmological Constant}},
  \href{http://dx.doi.org/10.1142/S0218271811020755,
  10.1007/s10714-011-1199-1}{\emph{Gen. Rel. Grav.} {\bfseries 43} (2011)
  2555--2560}, [\href{https://arxiv.org/abs/1105.3105}{{\ttfamily 1105.3105}}].

\bibitem{Casadio:2013uia}
R.~Casadio, \emph{{Lorentz invariance without trans-Planckian physics?}},
  \href{http://dx.doi.org/10.1016/j.physletb.2013.06.040}{\emph{Phys. Lett.}
  {\bfseries B724} (2013) 351--354},
  [\href{https://arxiv.org/abs/1303.1914}{{\ttfamily 1303.1914}}].

\bibitem{Kaloper:2013zca}
N.~Kaloper and A.~Padilla, \emph{{Sequestering the Standard Model Vacuum
  Energy}}, \href{http://dx.doi.org/10.1103/PhysRevLett.112.091304}{\emph{Phys.
  Rev. Lett.} {\bfseries 112} (2014) 091304},
  [\href{https://arxiv.org/abs/1309.6562}{{\ttfamily 1309.6562}}].

\bibitem{Kaloper:2015jra}
N.~Kaloper, A.~Padilla, D.~Stefanyszyn and G.~Zahariade, \emph{{Manifestly
  Local Theory of Vacuum Energy Sequestering}},
  \href{http://dx.doi.org/10.1103/PhysRevLett.116.051302}{\emph{Phys. Rev.
  Lett.} {\bfseries 116} (2016) 051302},
  [\href{https://arxiv.org/abs/1505.01492}{{\ttfamily 1505.01492}}].

\bibitem{Wang:2017oiy}
Q.~Wang, Z.~Zhu and W.~G. Unruh, \emph{{How the huge energy of quantum vacuum
  gravitates to drive the slow accelerating expansion of the Universe}},
  \href{http://dx.doi.org/10.1103/PhysRevD.95.103504}{\emph{Phys. Rev.}
  {\bfseries D95} (2017) 103504},
  [\href{https://arxiv.org/abs/1703.00543}{{\ttfamily 1703.00543}}].

\bibitem{Lombriser:2017cjy}
L.~Lombriser and V.~Smer-Barreto, \emph{{Is there another coincidence problem
  at the reionization epoch?}},
  \href{http://dx.doi.org/10.1103/PhysRevD.96.123505}{\emph{Phys. Rev.}
  {\bfseries D96} (2017) 123505},
  [\href{https://arxiv.org/abs/1707.03388}{{\ttfamily 1707.03388}}].

\bibitem{Appleby:2018yci}
S.~Appleby and E.~V. Linder, \emph{{The Well-Tempered Cosmological Constant}},
  \href{http://dx.doi.org/10.1088/1475-7516/2018/07/034}{\emph{JCAP} {\bfseries
  1807} (2018) 034}, [\href{https://arxiv.org/abs/1805.00470}{{\ttfamily
  1805.00470}}].

\bibitem{Lombriser:2018aru}
L.~Lombriser, \emph{{Late-time acceleration by a residual cosmological constant
  from sequestering vacuum energy in ultimate collapsed structures}},
  \href{https://arxiv.org/abs/1805.05918}{{\ttfamily 1805.05918}}.

\bibitem{Wang:2018kly}
S.-J. Wang, \emph{{Electroweak relaxation of cosmological hierarchy}},
  \href{http://dx.doi.org/10.1103/PhysRevD.99.023529}{\emph{Phys. Rev.}
  {\bfseries D99} (2019) 023529},
  [\href{https://arxiv.org/abs/1810.06445}{{\ttfamily 1810.06445}}].

\bibitem{Evnin:2018zeo}
O.~Evnin and K.~Nguyen, \emph{{Graceful exit for the cosmological constant
  damping scenario}},
  \href{http://dx.doi.org/10.1103/PhysRevD.98.124031}{\emph{Phys. Rev.}
  {\bfseries D98} (2018) 124031},
  [\href{https://arxiv.org/abs/1810.12336}{{\ttfamily 1810.12336}}].

\bibitem{Canales:2018tbn}
F.~Canales, B.~Koch, C.~Laporte and A.~Rincon, \emph{{Vacuum energy density:
  deflation during inflation}},
  \href{https://arxiv.org/abs/1812.10526}{{\ttfamily 1812.10526}}.

\bibitem{Lombriser:2019jia}
L.~Lombriser, \emph{{On the cosmological constant problem}},
  \href{https://arxiv.org/abs/1901.08588}{{\ttfamily 1901.08588}}.

\bibitem{Hildebrandt:2016iqg}
H.~Hildebrandt et~al., \emph{{KiDS-450: Cosmological parameter constraints from
  tomographic weak gravitational lensing}},
  \href{http://dx.doi.org/10.1093/mnras/stw2805}{\emph{Mon. Not. Roy. Astron.
  Soc.} {\bfseries 465} (2017) 1454},
  [\href{https://arxiv.org/abs/1606.05338}{{\ttfamily 1606.05338}}].

\bibitem{2018ApJ...855..136R}
A.~G. {Riess}, S.~{Casertano}, W.~{Yuan}, L.~{Macri}, J.~{Anderson}, J.~W.
  {MacKenty} et~al., \emph{{New Parallaxes of Galactic Cepheids from Spatially
  Scanning the Hubble Space Telescope: Implications for the Hubble Constant}},
  \href{http://dx.doi.org/10.3847/1538-4357/aaadb7}{\emph{\apj} {\bfseries 855}
  (Mar., 2018) 136}, [\href{https://arxiv.org/abs/1801.01120}{{\ttfamily
  1801.01120}}].

\end{thebibliography}\endgroup
%%%%%%%%%%%%%%%%%%%%%%%

\end{document}